\documentclass[16pt,superscriptaddress,showpacs,preprintnumbers,amsmath,amssymb,floatfix,prd]{revtex4}

\usepackage{graphicx}
\usepackage{dcolumn}
\usepackage{bm}
\usepackage{hyperref}
\hypersetup{colorlinks,citecolor=blue}
\usepackage{enumerate}

\begin{document}

\title{   Euclidean quantum wormholes  }

\author{Farook Rahaman}
\email{rahaman@associates.iucaa.in}
\affiliation{Department of Mathematics, Jadavpur University, Kolkata 700032, West Bengal, India}

\author{Bikramarka S Choudhury}
\email{bikramarka@gmail.com}
\affiliation{Department of Mathematics, Jadavpur University, Kolkata 700032, West Bengal, India}

\author{Anikul Islam}%
\email{ anikulislam0025@gmail.com  }
\affiliation{Department of Mathematics, Jadavpur University, Kolkata 700032, West Bengal, India}

\date{\today}

\begin{abstract}

We study wormhole  as the solution of the Wheeler-deWitt  (WdW )  equation  satisfying Hawking-Page  wormhole boundary conditions in Friedmann-Robertson-Walker (FRW) cosmology.  The quantum wormholes are formulated with arbitrary factor ordering of the Hamiltonian constraint operators with  perfect fluid  matter sources  as well as minimally coupled scalar fields.

\end{abstract}

\maketitle

\section{Introduction:}
 Usually there are two kinds of classical wormholes \cite{morris_throne_worm,einstein_rosen}:  Lorentzian and Euclidean.   Lorentzian wormholes \cite{Lorentzian_wormholes, reality_condi, Evolving_Lorentzian_wormhole} discussed earlier as   the vacuum solutions to the Lorentzian Einstein field equations, e.g.,
Schwarzschild wormholes \cite{schwarzschild_worm,schwarzschild_worm_trav},  Einstein-Rosen bridges \cite{einstein_rosen} etc.
On the other hand, Euclidean wormholes \cite{euclidean_quantum_grav} have been            considered        as    instantons,          solutions        of   the  classical
  Euclidean         Einstein        field equations,
which     consist      of  two    asymptotically flat    regions     connected        by   a  narrow throat (handle).
These     classical
wormholes        are  actually  saddle     points      of  the    Euclidean       action \cite{hilbert_action}     and    as a result       they     permit   the    Euclidean        path
integral       that can    be    approximated semiclassically.
Generally, these  wormholes \cite{bases_cosmology} are characterized by  quantum tunneling \cite{quant_tunnel, quantum_tunnelinhg} between
unalike zones of spacetime having mostly different topologies.
Researchers show interest in Euclidean wormhole physics after the discovery of the  solution  to  the  Einstein  equations
with a third-rank antisymmetric tensor field as the matter source which represents a wormhole or  bridge,
 joining   two  asymptotically  flat  regions  of  Euclidian  space. This shows that gravitational instantons \cite{grav_instantons}
could really exist. Thus it is interesting to know
 the feasible occurrence of wormhole solutions
arising
 through
the initial Euclidean era of the universe.
 Several research works    have been done on the physical consequences of wormhole dynamics and these studies can be categorized into three types.    First implies that  quantum coherence \cite{quantum_coherence} is not truly vanished because  wormholes join two asymptotically flat or de-Sitter regions by a throat of radius of the order of Planck length \cite{plank_length}.   Secondly,  microscopic wormholes  might offer the  mechanism that would resolve the cosmological constant problem \cite{cosmo_const_prob, cosmo_constant_problem}, whereas macroscopic wormholes are  responsible for the ultimate   evaporation and complete fading of black hole.  Third type is  terrible, which is, not all type of matter fields admit wormhole solutions.   However,  if wormholes regularize  important physical parameters, then   not only all kind of matter fields but also pure gravity should admit wormhole solutions.  This inspired   Hawking and Page \cite{hawking_page}  to interpret wormholes in a different way.
They suggested wormholes should be treated as solutions of the Wheeler-DeWitt (WdW ) equation \cite{wdw_eq,wdw_equation,A,B,C,D} under the suitable boundary conditions  instead of considering it as solutions to the classical field equations.
Here  the wave functional $\psi$ \cite{wave_function} must be regular and it  should be exponentially
damped for large three degenerated  geometry.  For the existence of wormhole, the boundary conditions   should be satisfied in minisuperspace models \cite{minisuperspace}  in stead of whole super-space. Researchers are interested whether all forms of matter fields really admit wormhole solution. In this paper, we will provide some discussions on the Einstein-Hilbert action with a minimally coupled scalar field \cite{minimally_coupled_scalar_field, minimally_scalar_field} and  perfect fluid  matter sources   to explore quantum and semiclassical wormhole solutions for some particular  form of the potentials as well as different equation of states  within the  background of FRW \cite{frw_model} minisuperspace model.

 \section{  Quantum wormhole with minimally coupled scalar fields:}

 Quantum wormhole    are non singular solutions
of the Wheeler - de Witt (WdW ) equation with specific  boundary conditions. Usually,
  the manifold is taken as  asymptotically Euclidean.
Hawking and Page have suggested that   the wave function for large 3-metrics should coincide  with the
 vacuum wave function defined by a path integral over all asymptotically
Euclidean metrics. It is argued that
 every matter or gauge field could admit wormhole solutions.  However this consideration is not true for all matter or gauge fields. Therefore,
 it is important  to search  which matter or gauge fields generate wormhole
solutions.

 Let us consider the gravitational action  of a   minimally coupled scalar field   coupled to gravity as
\begin{equation}~S = \int_M d^4x\sqrt{g} \left[ \frac{R}{16 \pi G} -\frac{1}{2\pi^2} \left(\frac{1}{2}\nabla_\mu \phi \nabla^\nu \phi + V(\phi)\right )\right]
- \frac{1}{8 \pi G}\int_\Sigma d^3x\sqrt{h}K, \end{equation}
where $V (\phi)$ is the potential of the scalar field $\phi$. h, K are defined as above.
Now  the metric for  a  FRW  universe is  taken to  be of the form
\begin{equation}~~~~~ds^2=- \epsilon dt^2 +a^2(t)\left[\frac{dr^2}{1-kr^2} + r^2 ( d\theta^2 + \sin^2 \theta d\varphi^2)\right].~~~~~\end{equation}
Here $\epsilon = \pm 1$ i.e. it  characterizes the signature change:    $\epsilon  =  1$  for  usual  Lorentzian  space and  $\epsilon  =  -1$
in  Euclidean  space and  the curvature parameter $k = 0, \pm 1$,  stands
for the flat, closed and open models respectively.
The action (1) can be written as
\begin{equation}~~~~~S = M \int \left [ -\frac{1}{2} a \dot a^2 + \frac{k a}{2} + \frac{1}{M}\left(\frac{1}{2}\dot \phi^2 -V(\phi)\right) a^3\right]dt .\end{equation}
 For this system, the Hamilton
constraint equation corresponding to (11.43) takes the following form
\begin{equation}~~~~~~ H = - \frac{1}{2Ma} P_a^2 + \frac{1}{2a^3} P_\phi^2 - \frac{M}{2}ka +a^3V ,~~~~~~~~\end{equation}
where we have chosen $ M  =  \frac{3\pi }{
2G}$
with  $G$  is Newton's constant.
Here,  $P_a $  and $P_\phi  $    are the corresponding momenta canonically conjugate to $a$ and $ \phi $  respectively.

Now  we try to write the most general form of WdW equation by using the hamiltonian operator acting  on the wave function {$\Psi$} in minisuperspace. According to Dirac quantization procedure we use the most general quantization of momenta   instead  of  $p_a=-i\hbar\frac{\partial}{\partial a} ~ and~ p_\phi=-i\hbar \frac{\partial}{\partial a}$ as
$$p^2_a=-\frac{\hbar^2}{a^p} \left(\frac{\partial}{\partial a}\left(a^p \frac{\partial}{\partial a}\right)\right)=-\hbar^2\left[\frac{\partial^2}{\partial a^2}+\frac{p}{a}\frac{\partial}{\partial a}\right],$$
 $$p^2_\phi=-\frac{\hbar^2}{\phi^q}\left(\frac{\partial}{\partial \phi}\left(a^p \frac{\partial}{\partial a}\right)\right)=-\hbar^2\left[\frac{\partial^2}{\partial \phi^2}+\frac{q}{\phi}\frac{\partial}{\partial \phi}\right].$$

Hence the WdW  equation $H\Psi=0$, in minisuperspace takes the following form by substituting the dynamical variables by the above operators as
\begin{equation}
\left[\frac{\hbar^2}{2M}\left(\frac{\partial^2}{\partial a^2}+\frac{p}{a}\frac{\partial}{\partial a}\right)-\frac{M}{2}ka^2-\frac{\hbar^2}{2a^2}\left(\frac{\partial^2}{\partial \phi^2}+\frac{q}{\phi}\frac{\partial}{\partial \phi}\right)+a^4 V(\phi)\right]\Psi=0  ,
\end{equation}
where p and q represents parts of the factor ordering ambiguities of the operator factors $a$   $\&$   $\frac{\partial}{\partial a}$ and $\phi$  $\&$ $\frac{\partial}{\partial \phi}$ respectively.
For different values of the ordering parameters p, q, one can get different mini super space models. It is known that quantum wormholes are  the solutions to the WdW equations which follow some boundary conditions. We follow Hawking-Page's proposals which are given below:
\begin{enumerate}[(i)]
    \item The wave function $\Psi$ is decaying with the scale factor as ${a \to \infty}$. This means the wormhole is asymptotically Euclidean.
    \item The wave function $\Psi$ is regular at $ a \to 0$.
    \item The wave function should be a convergent function. Due to the presence of matter, no divergent is occurred.
\end{enumerate}

The equation (5) involving potential $V(\phi)$ is a complicated equation, therefore,   we consider some simplifying
suppositions for potential to achieve wormhole solution.\\

\textbf{Case I :} $V(\phi)=0$:\\

 In this case (5) reads
 \begin{equation}
 \left[\frac{\hbar^2}{2M}\left(a^2 \frac{\partial^2}{\partial a^2}+pa\frac{\partial}{\partial a}\right)-\frac{M}{2}ka^4\right]\Psi(a,\phi)=\frac{\hbar^2}{2}\left(\frac{\partial^2}{\partial \phi^2}+\frac{q}{\phi}\frac{\partial}{\partial \phi}\right)\Psi(a,\phi)
 \end{equation}
 To solve this equation, one can use separation of variables as $\Psi(a,\phi)=A(a)B(\phi).$ Thus above equation (6) yields
 \begin{equation}
     \frac{1}{A}\left[\frac{\hbar^2}{2M}\left(a^2 \frac{\partial^2 A}{\partial a^2}+pa\frac{\partial A}{\partial a}\right)-\frac{M}{2}ka^4 A\right]=\frac{\hbar^2}{2B}\left(\frac{\partial^2 B}{\partial \phi^2}+\frac{q}{\phi}\frac{\partial B}{\partial \phi}\right)=\omega^2,
 \end{equation}
 where, $\omega^2$ is separation constant.

 Solving equation(7), one can get final solution of the wave function $\Psi(a,\phi)=A(a)B(\phi)$ for different values of $k, p, q$.

 \[A(a) = C_1a^{-\frac{p}{2} + \frac{1}{2} }  Bessel J \left(\frac{\sqrt{p^2 + 4 b_{00} - 2 p + 1}}{4}, \frac{\sqrt{-a_{00} k}a^2}{2}\right)\] \begin{equation}  + C_2a^{-\frac{p}{2}+ \frac{1}{2}} Bessel Y \left(\frac{\sqrt{p^2 + 4b_{00} - 2 p + 1}}{4}, \frac{\sqrt{-a_{00}k} a^2}{2}\right) ,\end{equation}
where $C_1 ~and ~C_2 $ are integration constants and  $ a_{00}= \frac{M^2}{\hbar^2} ~~~\& ~~b_{00}=\frac{2 M \omega^2}{\hbar^2}.$\\
Here, $BesselJ$ and $BesselY$ are the Bessel functions of first and second kind defined as \\
$BesselJ(v,x) := J_v (x) := \sum_{m=0}^{\infty} \frac{(-1)^m}{m! \Gamma(m+v+1)} {( \frac{x}{2} )}^{2m + v}$ and $BesselY(v,x) := Y_v (x) := \frac{J_v(x) cos(v \pi) - J_{-v}(x) }{sin(v \pi)}$

  \[B(\phi) = C_3a^{-\frac{q}{2} + \frac{1}{2} }  Bessel J \left(-\frac{q}{2} + \frac{1}{2}, \sqrt{c_{00} }\phi\right)\] \begin{equation}  + C_4a^{-\frac{q}{2} + \frac{1}{2} }  Bessel Y \left(-\frac{q}{2} + \frac{1}{2},  \sqrt{c_{00} }\phi\right) ,\end{equation}
where $C_3 ~and ~C_4 $ are integration constants and $ ~~c_{00}=\frac{2  \omega^2}{\hbar^2}.$

Now we will have to check whether the wave function satisfies the wormhole boundary conditions.  We have plotted the wave functions for different  values of the parameters  ( see figures 1-2 ) that represent the quantum wormhole solutions. Note that
the wave function $\Psi$ is decaying with the scale factor as ${a \to \infty}$. This indicates the wormhole is asymptotically Euclidean.
     Also, the wave function $\Psi$ is regular at $ a \to 0$ and is a
     convergent function.  \\

\begin{figure} [thbp]
\centering
	\includegraphics[width=7.6cm]{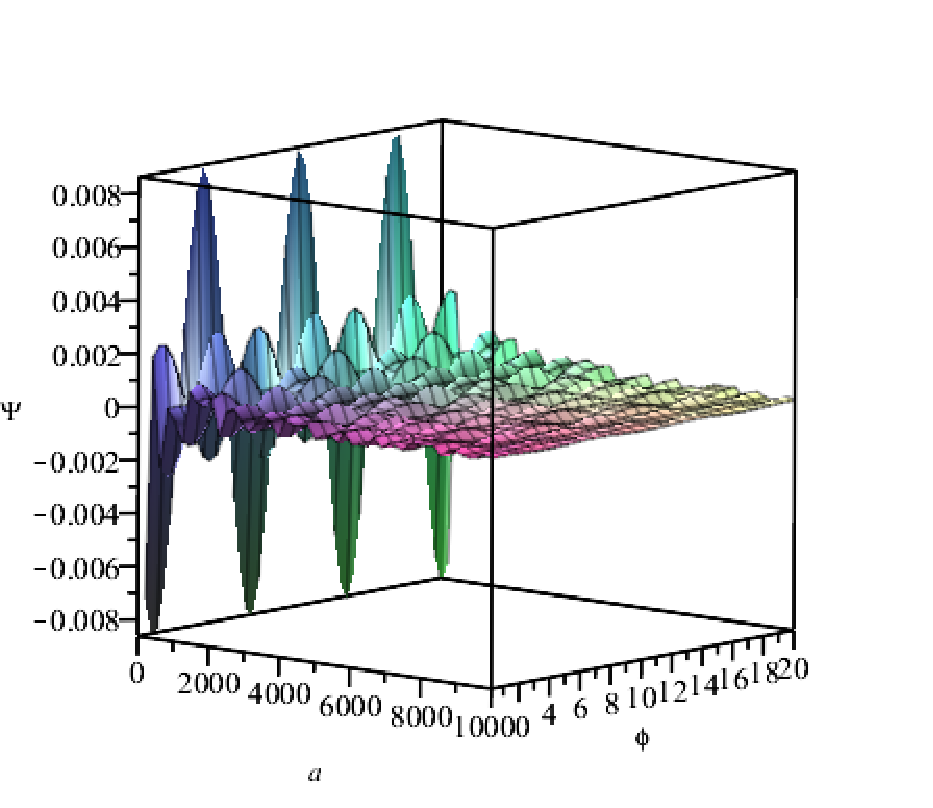}
\includegraphics[width=8cm]{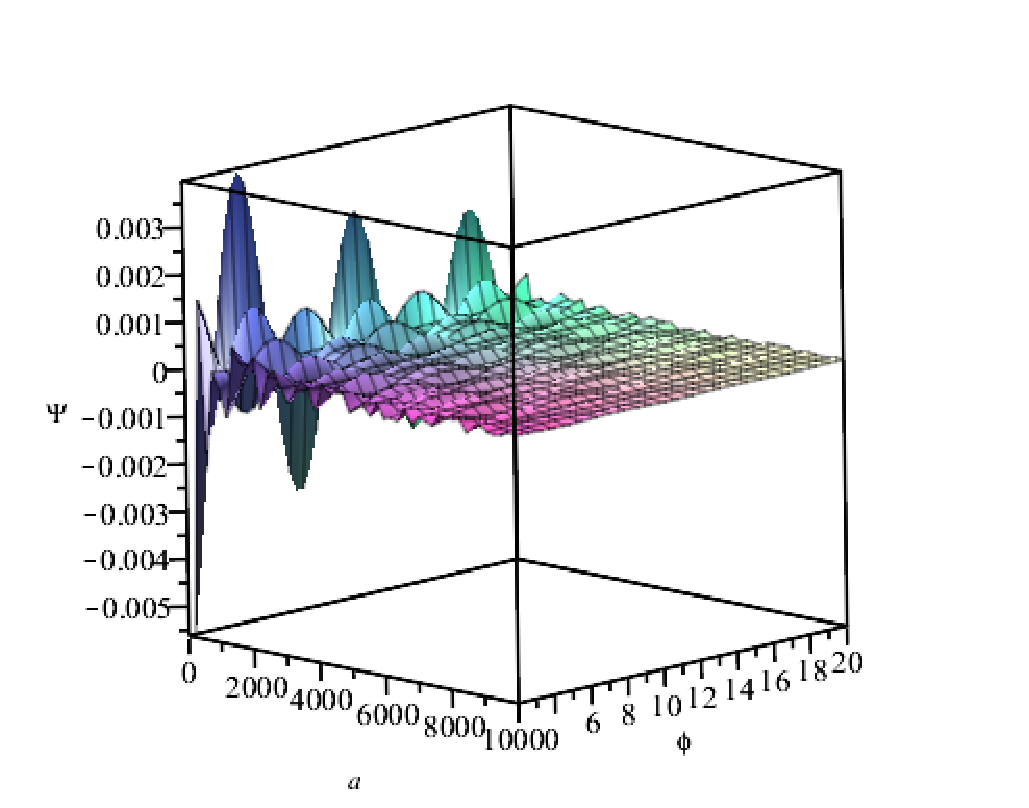}
	\caption  { The wave function represents the quantum wormhole solution (left panel) for $p=1, k= -1, q=0$,    (right panel) for $p=1, k= -1, q=1$  .}
\end{figure}
\begin{figure} [thbp]
\centering
	\includegraphics[width=7cm]{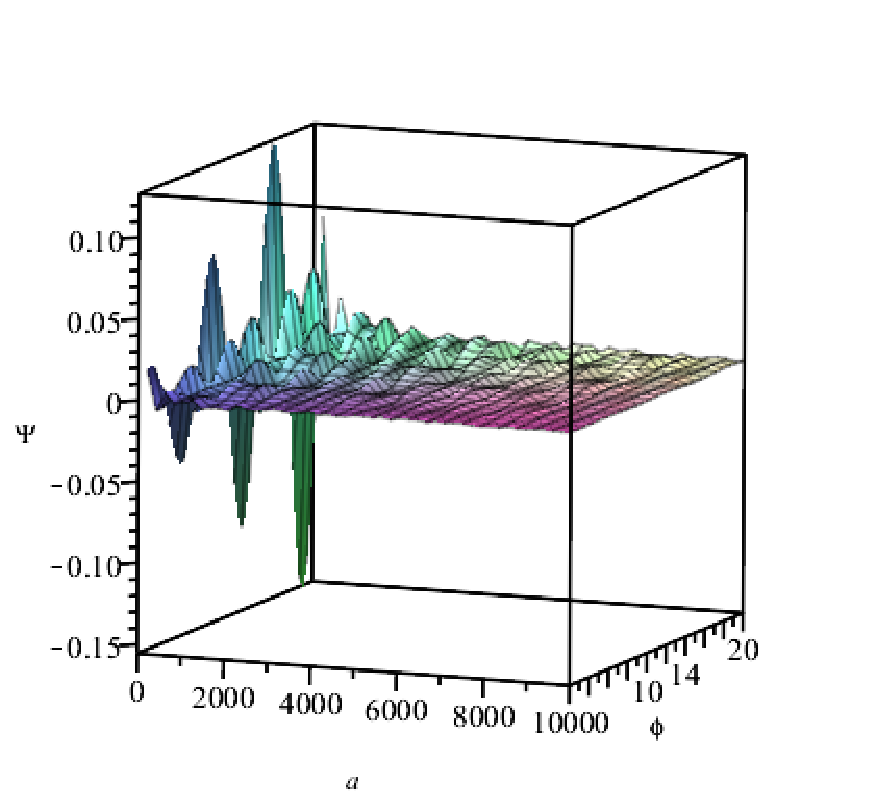}
\includegraphics[width=8cm]{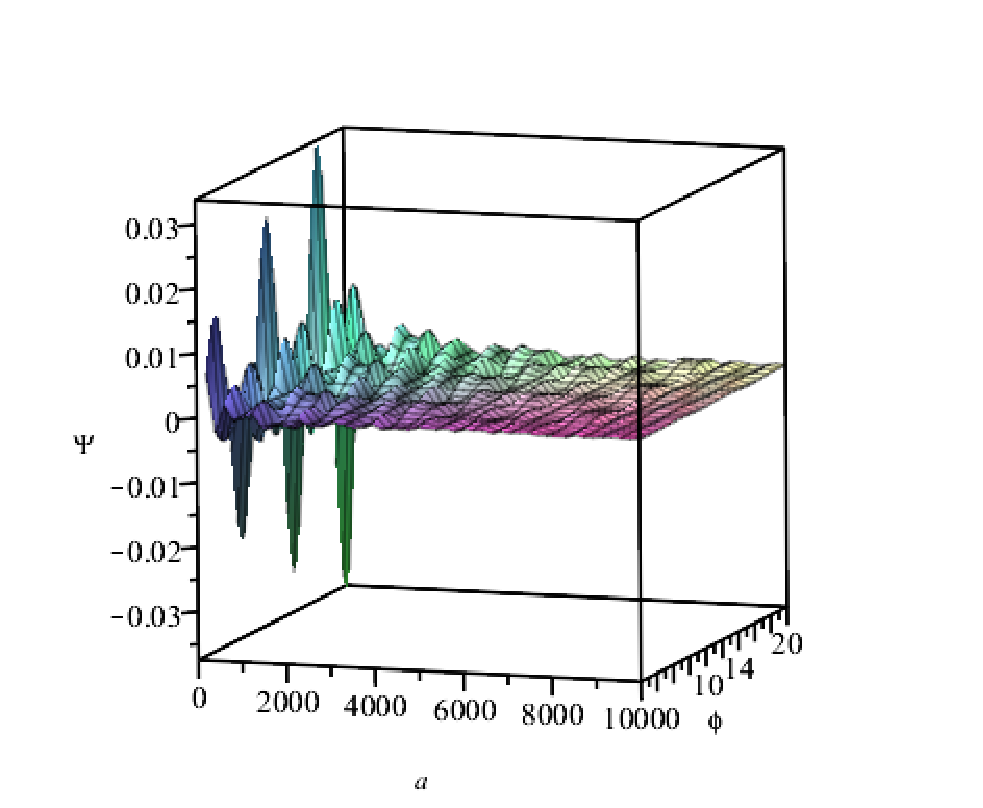}
	\caption { The wave function represents the quantum wormhole solution (left panel) for $p=1, k= -1, q=-2$,       (right panel) for $p=1, k= -1, q=-1$ .}
\end{figure}

  \textbf{CASE II :} $V(\phi)=V_0$, where $V_0$ is a constant.\\

  Now WDW equation (5) assumes the form as

\begin{equation}
 \left[\frac{\hbar^2}{2M}\left(a^2 \frac{\partial^2}{\partial a^2}+pa\frac{\partial}{\partial a}\right)-\frac{M}{2}ka^4+a^6V_0\right]\Psi(a,\phi)=\frac{\hbar^2}{2}\left(\frac{\partial^2}{\partial \phi^2}+\frac{q}{\phi}\frac{\partial}{\partial \phi}\right)\Psi(a,\phi)
 \end{equation}

As before, using separable form of $\Psi(a,\phi)=A(a)B(\phi)$, one can get

\begin{equation}
     \frac{1}{A}\left[\frac{\hbar^2}{2M}\left(a^2 \frac{\partial^2 A}{\partial a^2}+pa\frac{\partial A}{\partial a}\right)-\frac{M}{2}ka^4 A+a^6V_0 A\right]=\frac{\hbar^2}{2B}\left(\frac{\partial^2 B}{\partial \phi^2}+\frac{q}{\phi}\frac{\partial B}{\partial \phi}\right)=\omega^2 ,
 \end{equation}
where, $\omega^2$ is separation constant.\\

 These equations yield solutions for A and B for different values of $k, p, q$. The equation containing $a$ is a very complicated equation, so we will discuss the behaviors of the wave function $\psi$ near $a \approx 0$ and $a \rightarrow \infty$. In the limit $a \rightarrow 0$, the terms $a^4 , ~ a^6$ are much smaller than unity whereas in the limit $a \rightarrow \infty$  , the terms $ a^4$ and constant are  smaller compared to the term   $a^6$. Therefore,    we
only keep the constant term in the first case and the term  $a^6$ for the latter case. Now the equation (11) assumes the following forms
\[\left(a^2 \frac{\partial^2 A}{\partial a^2}+pa\frac{\partial A}{\partial a}\right) +c_{11} A =0,\]
\[\left(a^2 \frac{\partial^2 A}{\partial a^2}+pa\frac{\partial A}{\partial a}\right) +b_{11} a^6A =0,\]
where, $c_{11} = \frac{2M \omega^2}{\hbar^2}$ and $b_{11} = \frac{2M V_0}{\hbar^2}$. The solutions of the above equations are found as
\begin{equation}A(a) = C_1a^{-\frac{p}{2 } + \frac{1}{2} + \frac{\sqrt{p^2 - 4c_{11 } - 2p + 1}}{2}} + C_2a^{-\frac{p}{2 } + \frac{1}{2} - \frac{\sqrt{p^2 - 4c_{11 } - 2p + 1}}{2}}, \end{equation}
\begin{equation}A(a) = C_3a^{-\frac{p}{2 } + \frac{1}{2} }Bessel J\left( \frac{p}{6 } - \frac{1}{6}, \frac{\sqrt {b_{11}}a^3}{3}\right) + C_4a^{-\frac{p}{2 } + \frac{1}{2} }Bessel Y \left( \frac{p}{6 } - \frac{1}{6}, \frac{\sqrt {b_{11}}a^3}{3}\right), \end{equation}
where $C_i$ are integration constants.

The    $\phi$ part has the same form as given in equation (9).
The behavior of the wave functions are shown in figures (3-4).\\

\begin{figure} [thbp]
\centering
	\includegraphics[width=8.5cm]{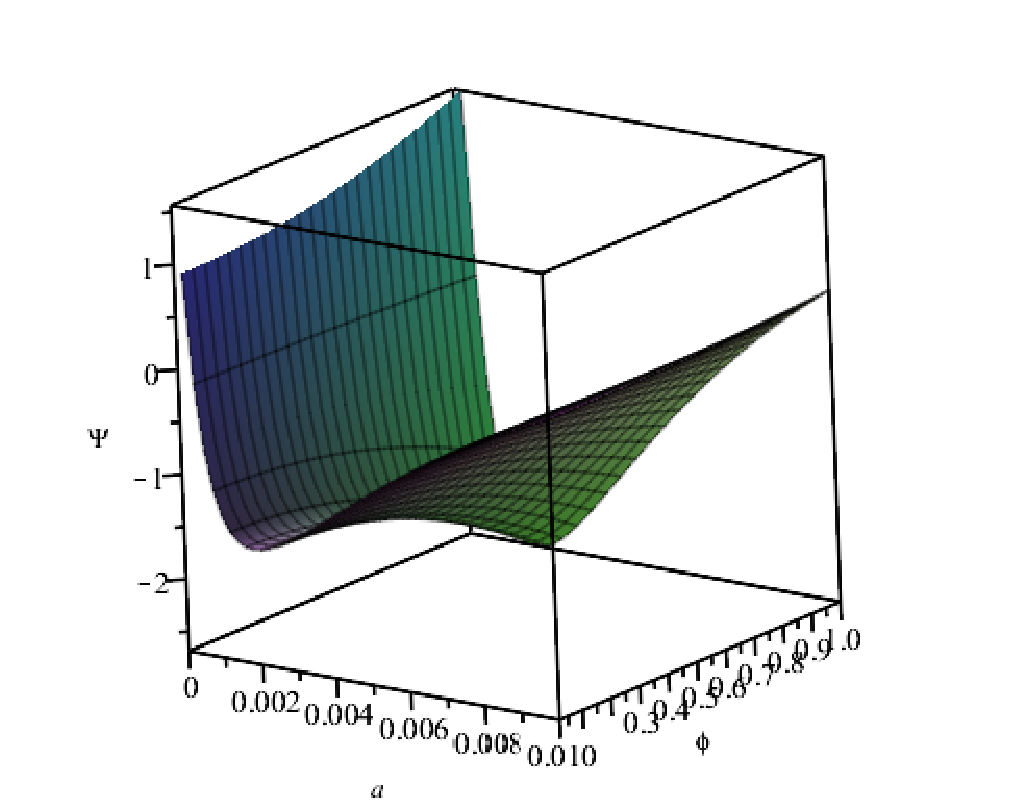}
\includegraphics[width=8cm]{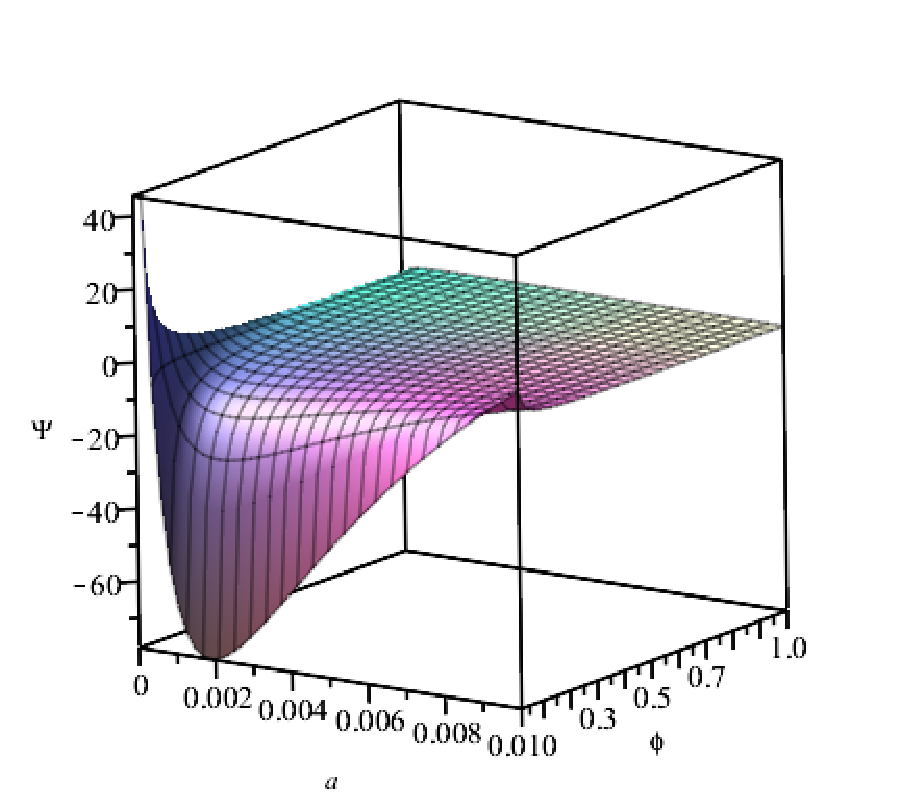}
	\caption { The wave function for scale factor $a \approx 0$ represents the quantum wormhole solution (left panel) for $p=1,  q=-2$,       (right panel) for $p=1,  q=2$ .}
\end{figure}
\begin{figure} [thbp]
\centering
	\includegraphics[width=8.5cm]{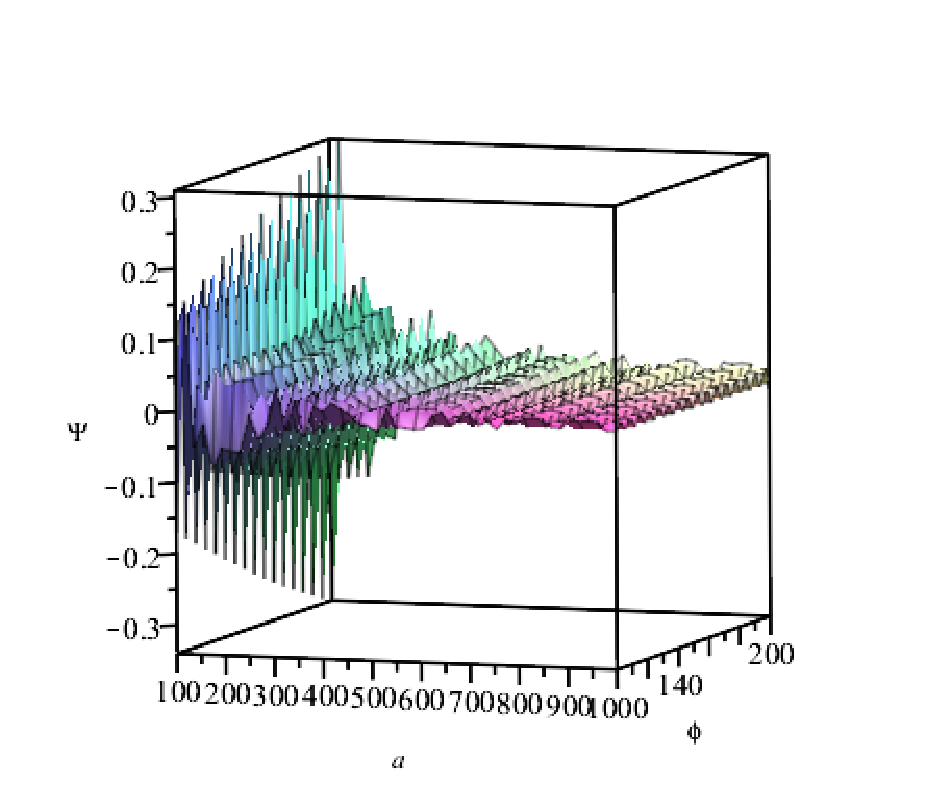}
\includegraphics[width=8.5cm]{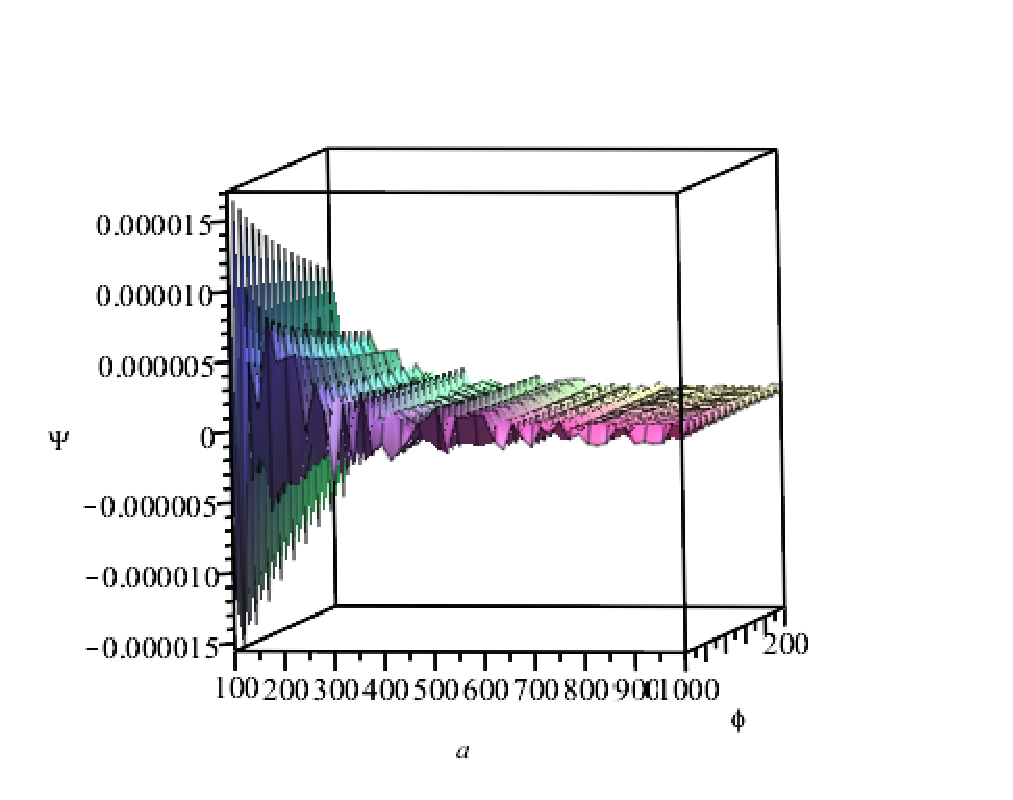}
	\caption {  The wave function for large value of scale factor i.e. $a \rightarrow \infty$,  represents the quantum wormhole solution (left panel) for $p=1,  q=-2$,       (right panel) for $p=1,  q=2$ .}
\end{figure}

 \textbf{CASE III :}  $V(\phi)$ is some function of $a$ i.e.  $V(\phi)=g(a)$.\\

In this case,  $V(\phi)$ is a function of scale factor $a$, therefore,  WdD equation (5) takes the form (after using separable form of $\Psi(a,\phi)=A(a)B(\phi)$,
 \begin{equation}
     \frac{1}{A}\left[\frac{\hbar^2}{2M}\left(a^2 \frac{\partial^2 A}{\partial a^2}+pa\frac{\partial A}{\partial a}\right)-\frac{M}{2}ka^4 A+a^6g(a)A\right]=\frac{\hbar^2}{2B}\left(\frac{\partial^2 B}{\partial \phi^2}+\frac{q}{\phi}\frac{\partial B}{\partial \phi}\right)=\omega^2  ,
 \end{equation}
 where, $\omega^2$ is separation constant.\\
We will solve this equation ( $a$ part ) by assuming the polynomial form of $g(a)$  as $g(a) = a^n$ , where n is an arbitrary constant.

We have found the solutions of $A(a)$ for the following cases:\\

\textbf{case-1: $n=-2$}
\[ A(a) = C_1 a^{-\frac{p}{2} + \frac{1}{2}} BesselJ \left(  \frac{\sqrt{p^2 - 4 c_{22} - 2p +1}}{4} , \frac{\sqrt{-a_{22} k + b_{22}}a^2}{2}  \right) \]\begin{equation} + C_2 a^{-\frac{p}{2} + \frac{1}{2}} BesselY \left(  \frac{\sqrt{p^2 - 4 c_{22} - 2p +1}}{4} , \frac{\sqrt{-a_{22} k + b_{22}}a^2}{2}. \right),\end{equation}

\textbf{case-2: $n=-6$}
  \[ A(a) = C_3 a^{-\frac{p}{2} + \frac{1}{2}} BesselJ \left(  \frac{\sqrt{p^2 - 4 b_{22} - 4 c_{22} - 2p +1}}{4} , \frac{\sqrt{-a_{22} k}\,a^2}{2}  \right) \] \begin{equation} +\,  C_4 a^{-\frac{p}{2} + \frac{1}{2}} BesselY \left(  \frac{\sqrt{p^2 - 4 b_{22} - 4 c_{22} - 2p +1}}{4} , \frac{\sqrt{-a_{22} k}\,a^2}{2} \right),
   \end{equation}
where  $ a_{22}= \frac{M^2}{\hbar^2} ~~~\& ~~b_{22}=\frac{2 M \omega^2}{\hbar^2}$, $ ~~c_{22}=\frac{2 M \omega^2}{\hbar^2},$
 and  $C_i$ are integration constants. The behavior of the wave functions are shown in figures (5-6).\\

\begin{figure} [thbp]
\centering
	\includegraphics[width=8.5cm]{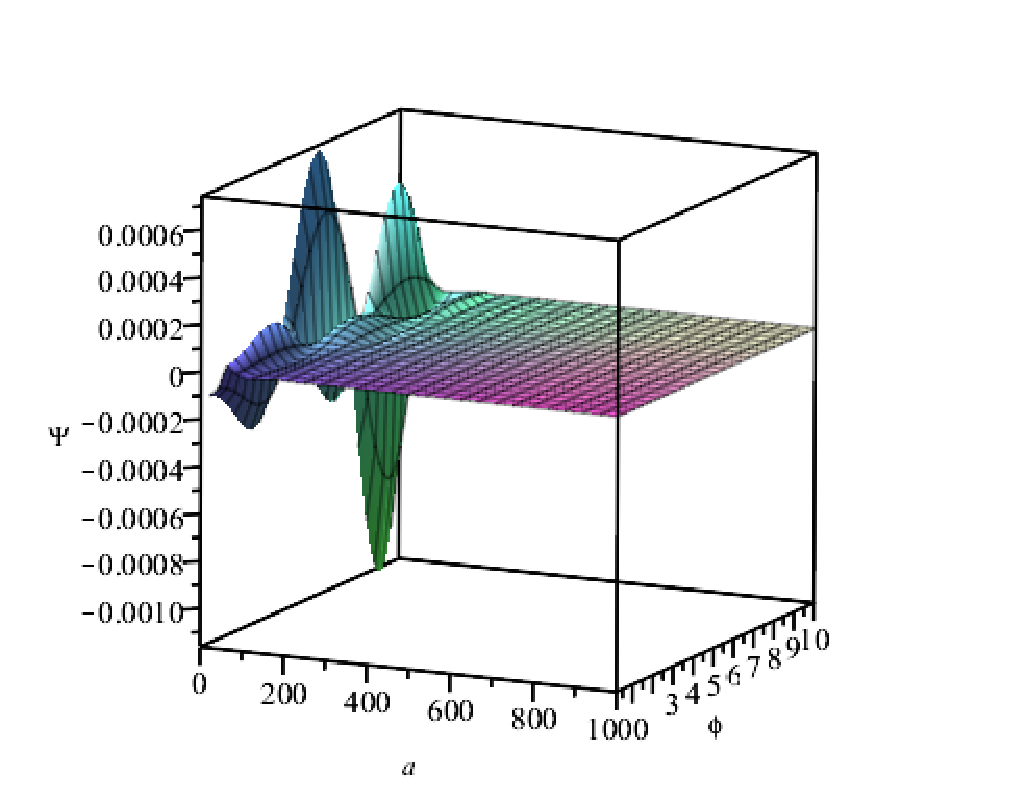}
\includegraphics[width=8cm]{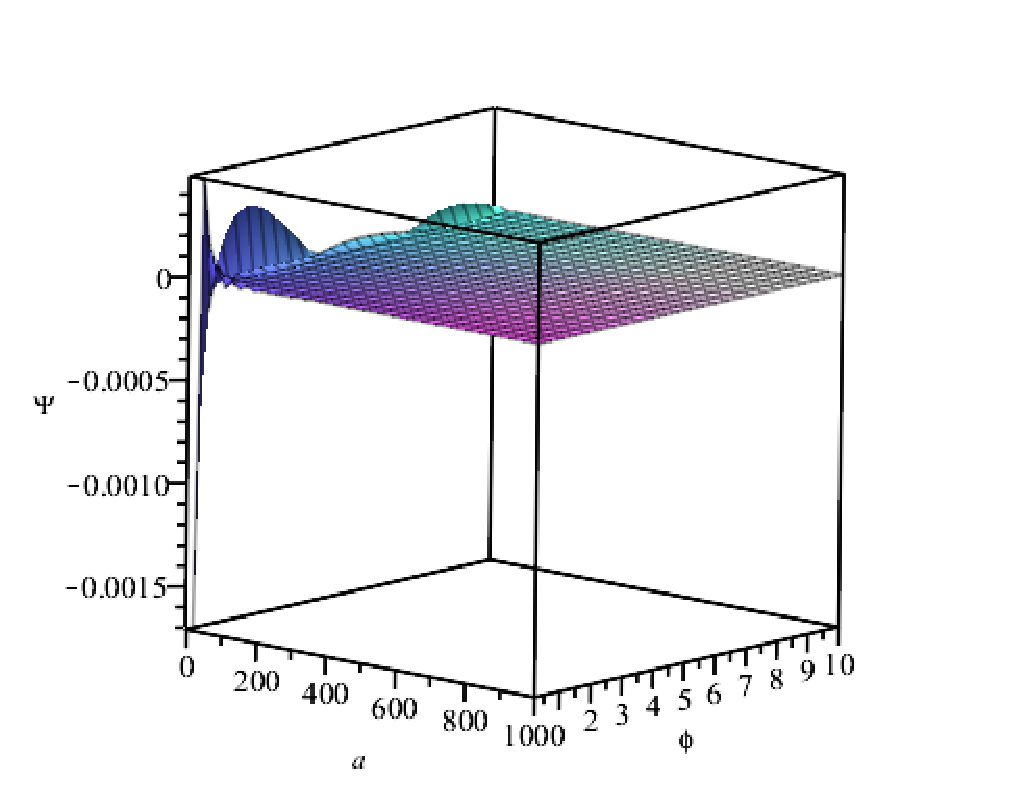}
	\caption {  The wave function represents the quantum wormhole solution (left panel) for $p=4,  q=-2, n=-2, k=-1$,       (right panel) for $p=4,  q=2, n=-2, k=-1$.}
\end{figure}
\begin{figure} [thbp]
\centering
	\includegraphics[width=8.5cm]{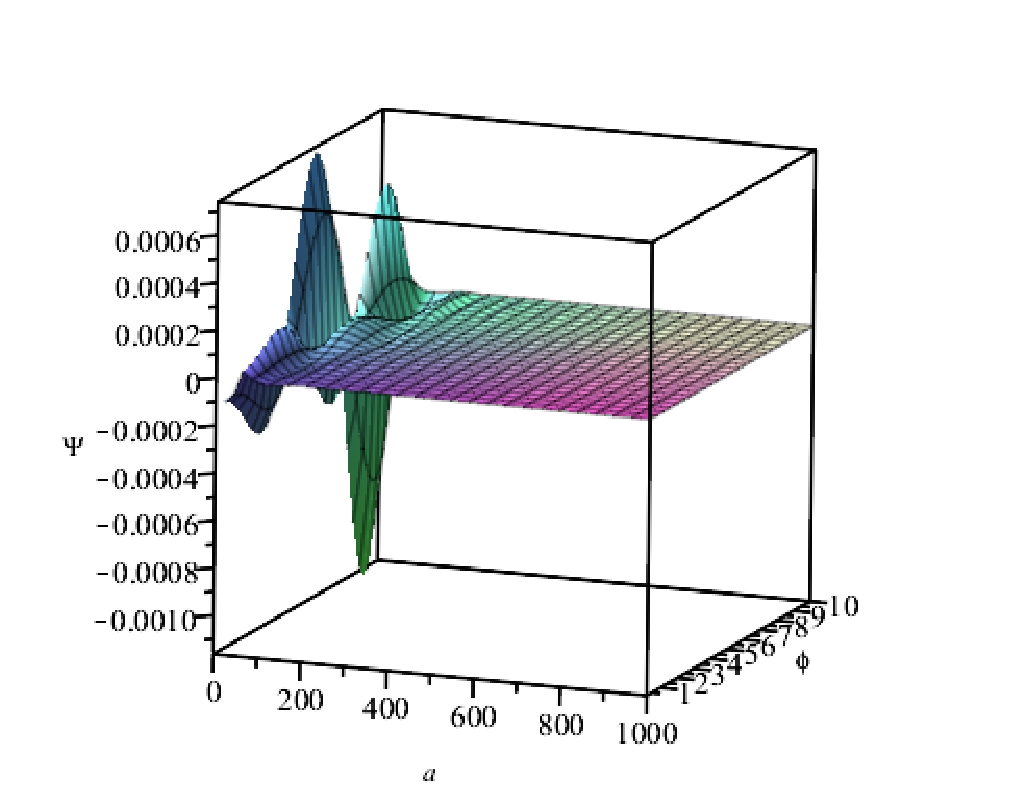}
\includegraphics[width=8.5cm]{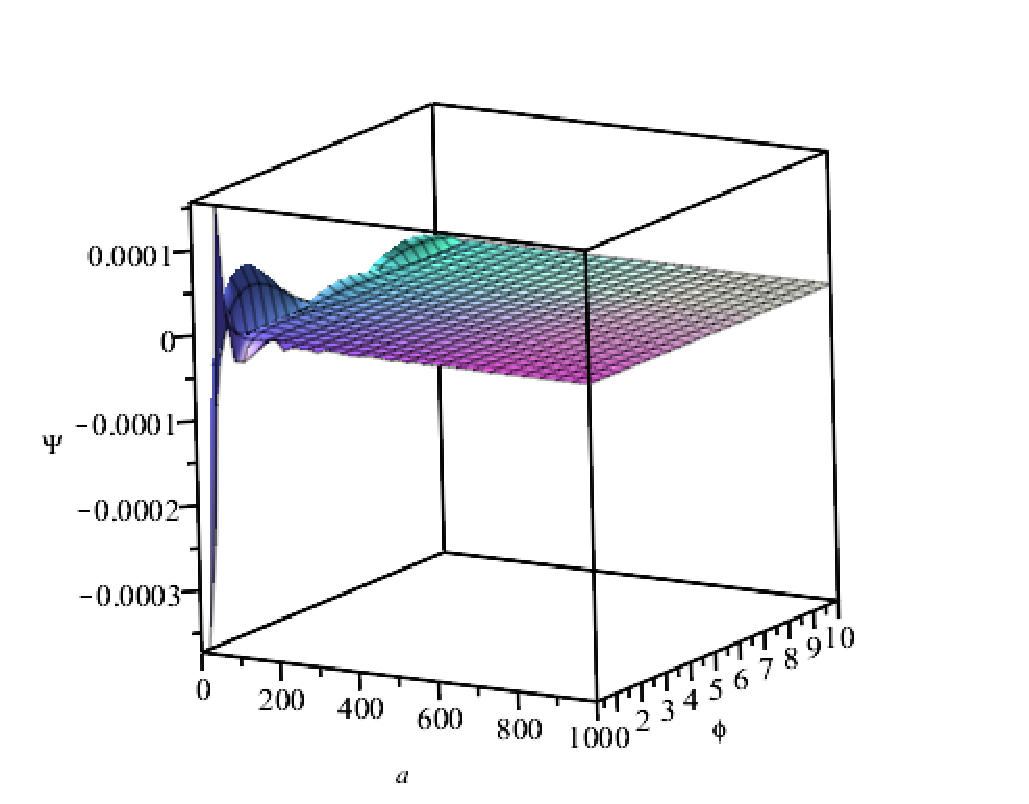}
	\caption {   The wave function represents the quantum wormhole solution (left panel) for  $p=4,  q=-2, n=-6, k=-1$,       (right panel) for for $p=4,  q=2, n=-6, k=-1$.}
\end{figure}

 \textbf{CASE IV :} $V=V_0\phi^{-\alpha}$, $V_0$ and $\alpha$ (may be positive or negative) are constants.\\

 For this power law form of the potential, we use the following transformation
 \begin{equation}
     \eta=a^m \phi^n ,
 \end{equation}
where $m,n$  are constants. To obtain quantum wormhole with this transformation (17), the WdW equation (5) takes the form as
\begin{equation}
    \frac{\hbar^2}{2M}\left(a^2\Psi_{aa}+ap\Psi_a\right)-\frac{M}{2}ka^4\Psi=\eta^{\frac{2n-2}{n}}a^{\frac{2m}{n}}\left[\frac{\hbar^2}{2}\left(n^2\Psi_{\eta \eta}+\frac{n(n-1)}{\eta}\psi_{\eta}+nqa^\frac{4m}{n}\eta^{\frac{3n-4}{n}}\Psi_\eta\right)-V_0a^{(6+\frac{m(\alpha-2)}{n})}\eta^{\frac{2-2n-\alpha}{n}}\Psi\right] .\end{equation}

Here the separation of variable is possible only for $m=\frac{6n}{2-\alpha}$ with $\alpha\neq2$ and $q=0$ to get a solution of the wave function of the form $\Psi=A(a)B(\eta)$. Hence we have
\begin{equation}
    \frac{a^{-\frac{2m}{n}}}{A}\left[\frac{\hbar^2}{2M}\left(a^2A_{aa}+apA_a\right)-\frac{M}{2}ka^4A\right]=\frac{\eta^{\frac{2n-2}{n}}}{B}\left[\frac{\hbar^2}{2}\left(n^2B_{\eta \eta}+\frac{n(n-1)}{\eta}B_\eta\right)-V_0\eta^{\frac{2(3-m)}{m}}B\right]=\omega^2 ,
\end{equation}
where, $\omega^2$ is separation constant.\\\\

 We have found the solutions of $A(a)$ for the following cases:\\

\textbf{case - 1:}  $\frac{2m}{n} = 2$:
   \[ A(a) = C_1 e^{-\frac{\sqrt{a_{33} k} \, a^2}{2} } KummerM \left( \frac{\sqrt{a_{33} k} (p+1) - b_{33} }{4 \sqrt{a_{33} k}} , \frac{p}{2} + \frac{1}{2}, \sqrt{a_{33} k} a^2 \right)      \]\begin{equation}
       + \, C_2 e^{-\frac{\sqrt{a_{33} k} \, a^2}{2} } KummerU \left( \frac{\sqrt{a_{33} k} (p+1) - b_{33} }{4 \sqrt{a_{33} k}} , \frac{p}{2} + \frac{1}{2}, \sqrt{a_{33} k} a^2 \right)
   \end{equation}

where $KummerM(a,b,x)$ is the Kummer's function $M(a,b,x)$ and $KummerU(a,b,x)$ is the Tricomi's function $U(a,b,x)$ both of which are solutions of Kummer's differential equation.
\\
\\
\textbf{case - 2:} $\frac{2m}{n} = 4$:
\begin{equation}
   A(a) = C_3 a^{- \frac{p}{2} + \frac{1}{2}} BesselJ \left( \frac{p}{4} - \frac{1}{4} , \frac{\sqrt{- a_{33} k + b_{33}}\, a^2 }{2} \right)
       + \,  C_4 a^{- \frac{p}{2} + \frac{1}{2}} BesselY \left( \frac{p}{4} - \frac{1}{4} , \frac{\sqrt{- a_{33} k + b_{33}}\, a^2 }{2} \right)
   \end{equation}
where $a_{33} = \frac{M^2}{\hbar^2}$, $b_{33} = \frac{2 \omega^2 M}{\hbar^2}$ and $C_i$ are integration constants.\\\\

 Exact solutions of $B(\eta)$ can be found for the following cases:\\

\textbf{case-1: $n=1, m=1$}
\[B(\eta) = C_1 \exp\left(-\frac{\eta^3}{3}\right) Heun T\left(\frac{(12)^{\frac{2}{3}}}{4}, 0, 0, \frac{3^{\frac{2}{3}}2^{\frac{1}{3}}\eta}{3}\right )\]\[+ C_2\exp\left(-\frac{\eta^3}{3}\right) Heun T\left(\frac{12^{\frac{2}{3}}}{4}, 0, 0, \frac{3^{\frac{2}{3}}2^{\frac{1}{3}}\eta}{3}\right)\int\frac{\exp \left(\frac{2\eta^3}{3}\right)}{HeunT \left (\frac{(12)^{\frac{2}{3}}}{4}, 0, 0, \frac{3^{\frac{2}{3}}2^{\frac{1}{3}}\eta}{3}\right )^2} d \eta),\]

where, $HeunT(a,b,c,x)$ is the Heun triconfluent function
\\
\\

\textbf{case-2: $n=1, m=2$}
\[B(\eta) =C_3 Airy Ai (\eta - 1) + C_4 Airy Bi(\eta - 1),\]
where $C_i$ are integration constants and here we have assumed  $V_0$ =1, and $AiryAi(x)$ and $AiryBi(x)$ are the linearly independent solutions to the airy wave differential equation $\frac{d^2 y}{d x^2} - xy = 0$
\\
\\

\begin{figure} [thbp]
\centering
	\includegraphics[width=5cm]{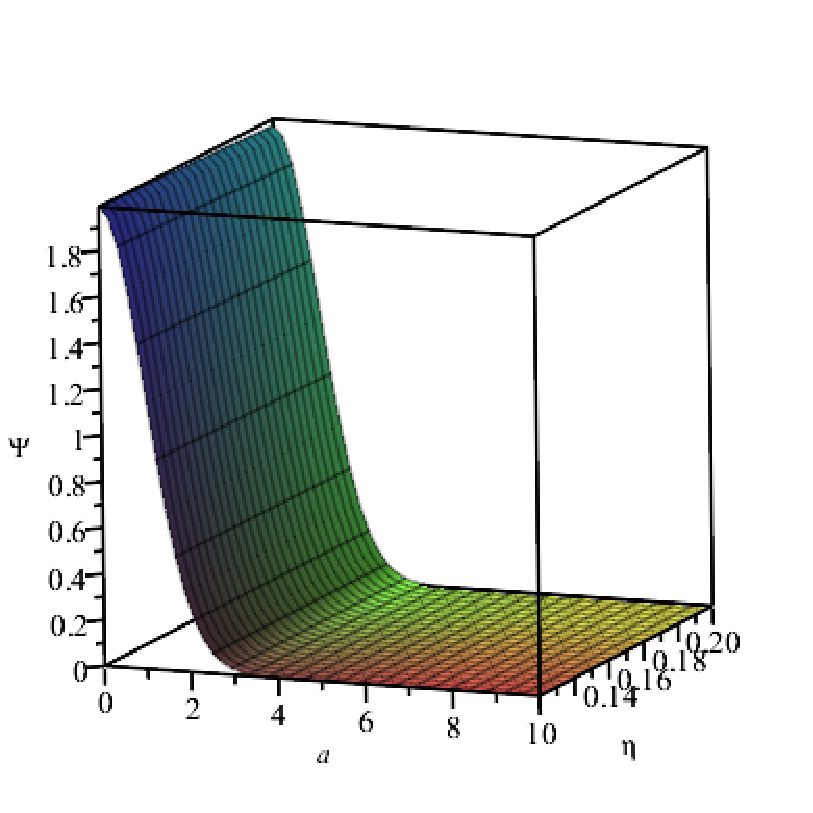}
\includegraphics[width=6.3cm]{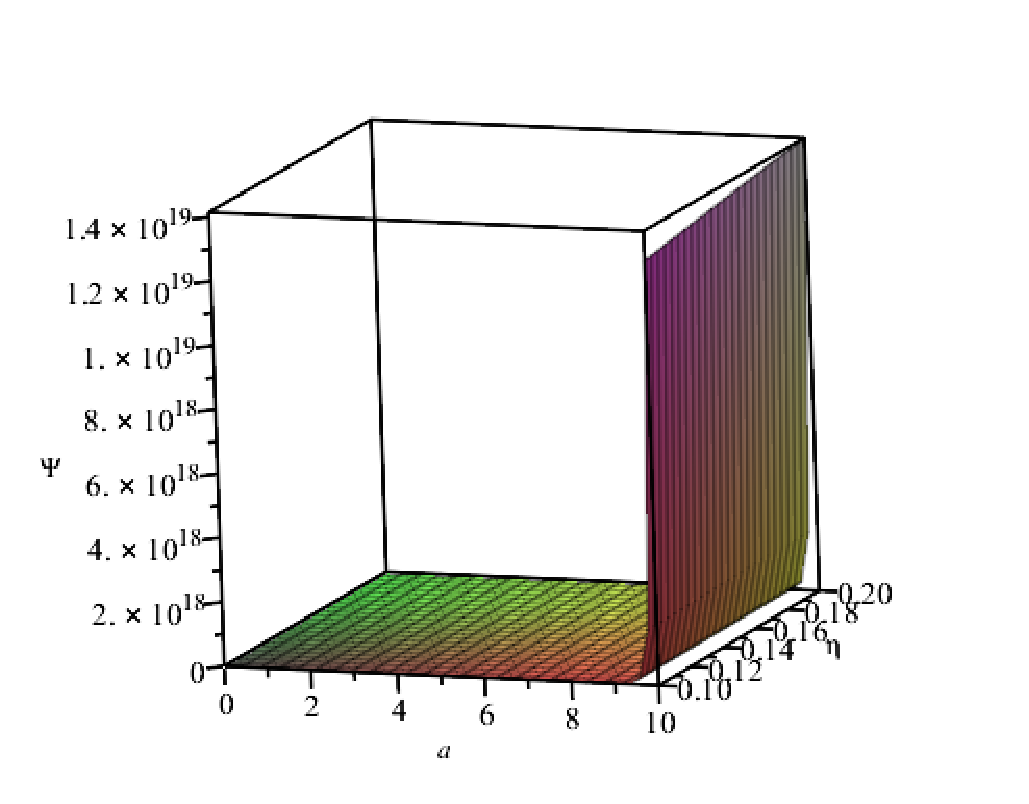}
\includegraphics[width=6.3cm]{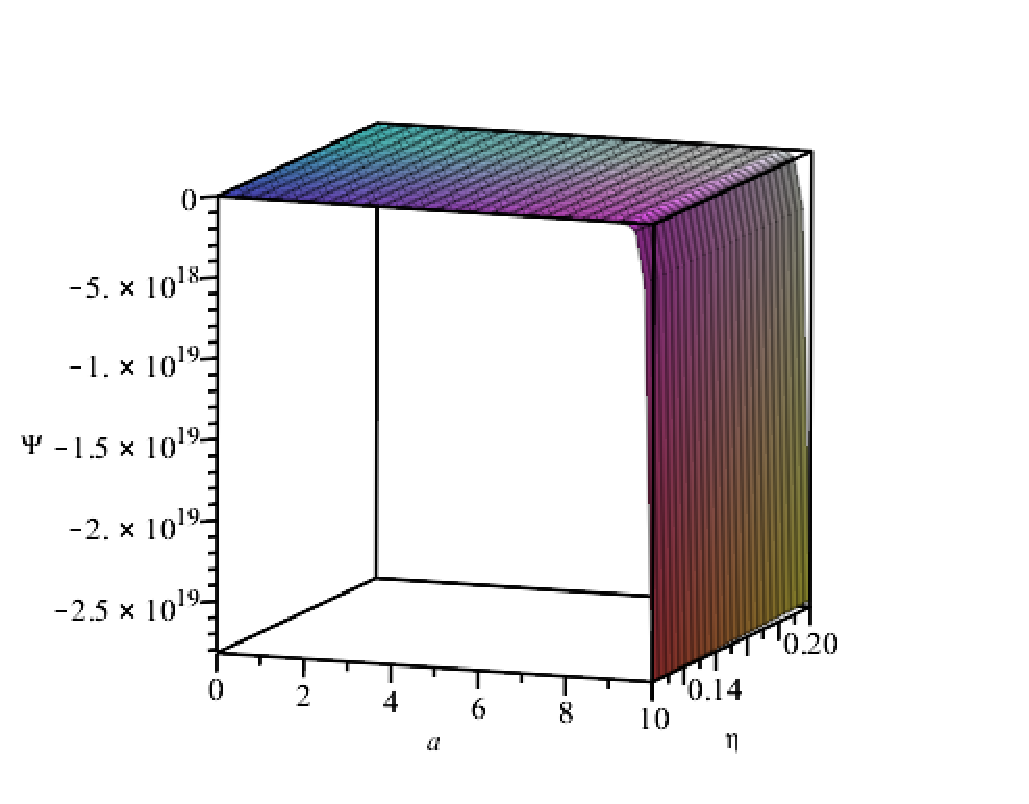}
	\caption {  The wave function for different values of the parameters.   (left panel) for $m=1,  n=1, p=0, k=1$,    (middle  panel) for $m=1,  n=1, p=-0.1, k=1$,   (right panel) for $m=1,  n=1, p=0.1, k=1$. Only figure in left panel represents a quantum wormhole solution. Other figures do not satisfy the viable wormhole conditions. }
\end{figure}

\begin{figure} [thbp]
\centering
	\includegraphics[width=5cm]{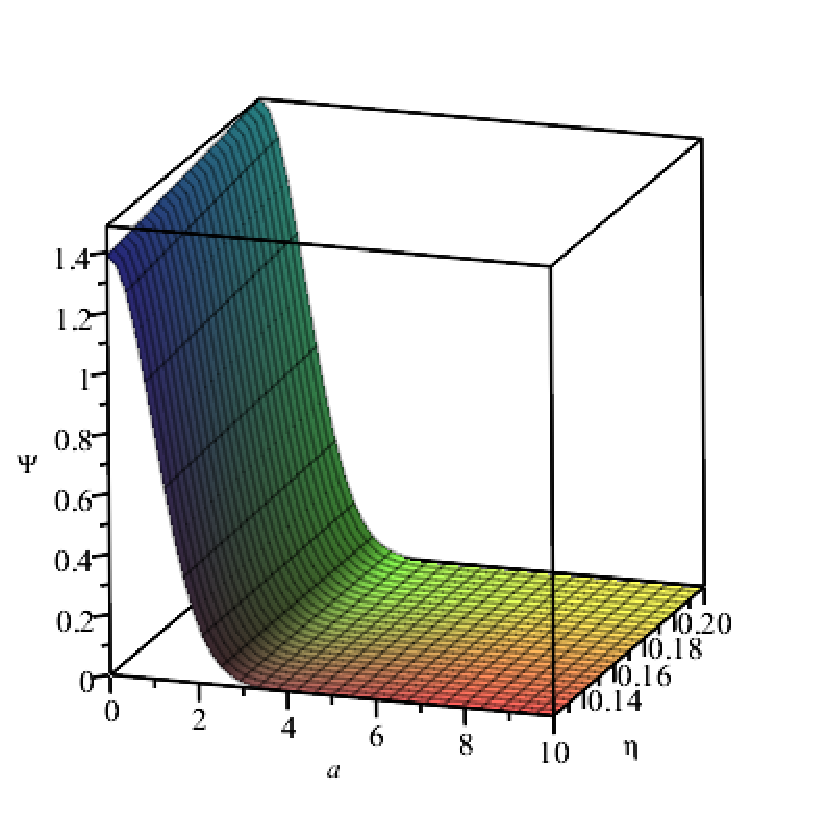}
\includegraphics[width=6.3cm]{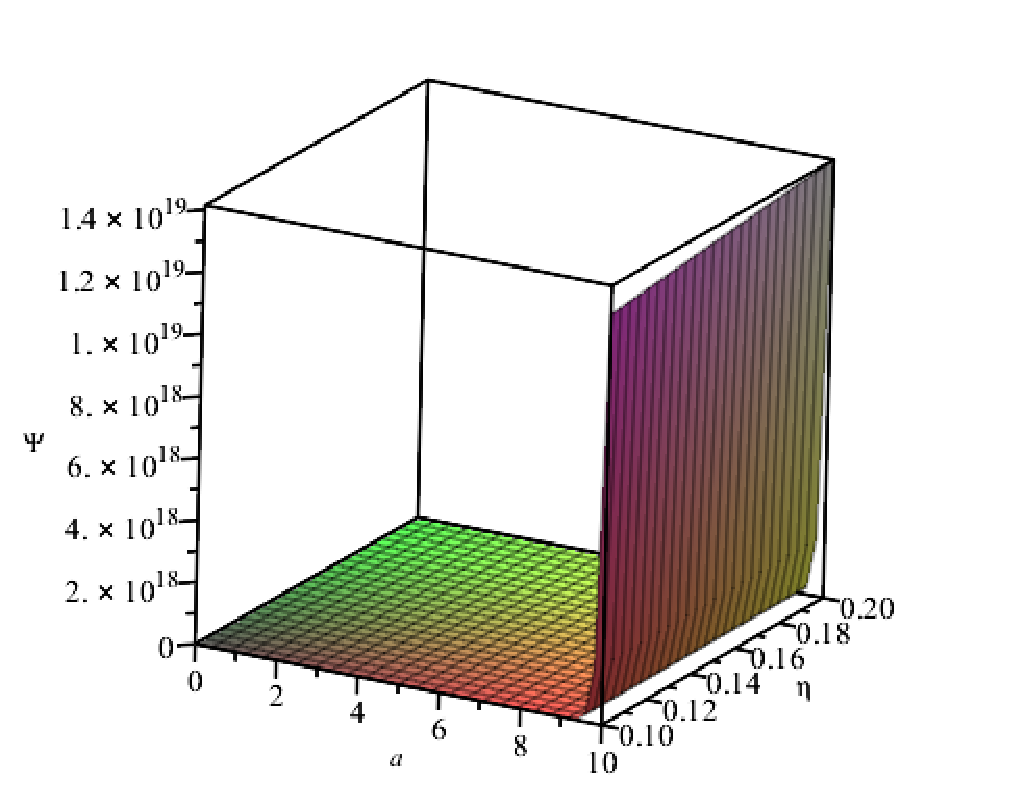}
\includegraphics[width=6.3cm]{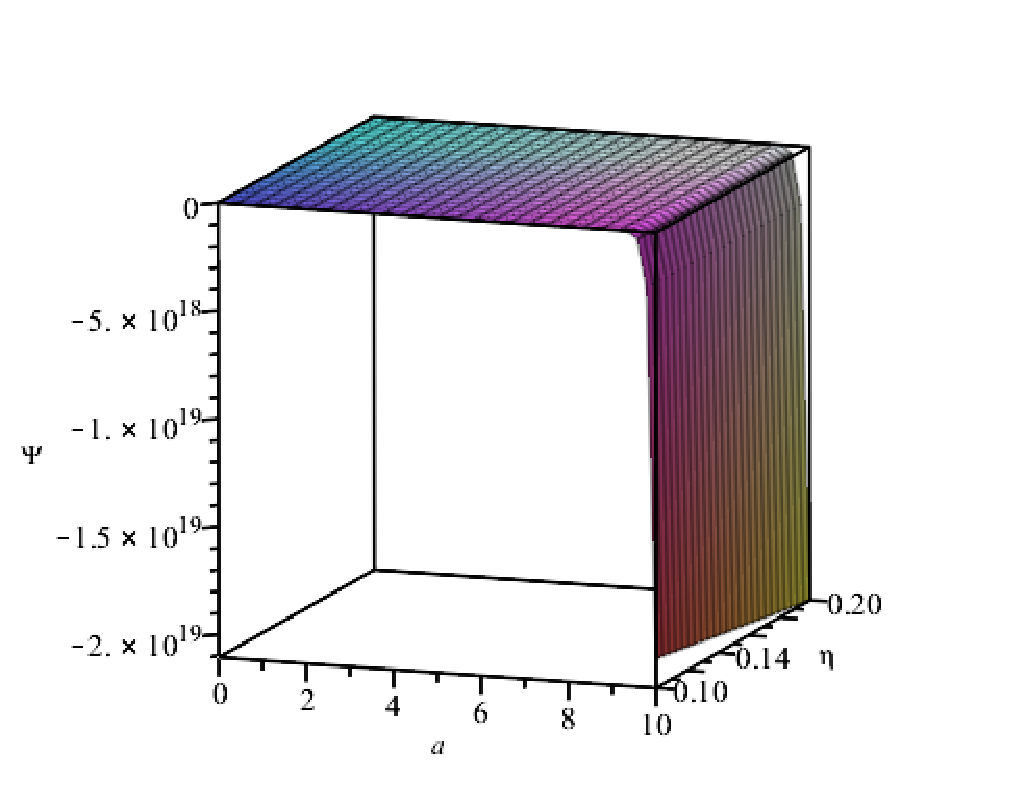}
	\caption { The wave function for different values of the parameters.   (left panel) for $m=2,  n=1, p=0, k=1$,    (middle  panel) for $m=2,  n=1, p=-0.1, k=1$,   (right panel) for $m=2,  n=1, p=0.1, k=1$. Only figure in left panel represents a quantum wormhole solution. Other figures do not satisfy the viable wormhole conditions. }
\end{figure}

\textbf{CASE V :} $V=V_0 e^{-\frac{\phi}{\lambda}}$, where $V_0$, $\lambda$ are constants. \\

For this exponential form
of potential, we can solve the WdW equation (5) by the method of separation of variables $\Psi(a,\phi)=A(a)B(y)$ by making a change of variables as $y=a^6e^{-\frac{\phi}{\lambda}}$. Now the WdW equation (5) takes the form
\begin{equation}
    \frac{1}{A}\left[\frac{\hbar^2}{2M}\left(a^2A_{aa}+apA_a\right)-\frac{M}{2}ka^4A\right]=\frac{1}{B}\left[\frac{\hbar^2}{2\lambda^2}\left(y^2B_{yy}+yB_y-\frac{yq}{\lambda\phi}B\right)-V_0yB\right]=\omega^2 .\end{equation}

Note that separation of variables is possible only when $q=0$. So for $q=0$, one can have two equations involving A and B separately. The solutions for    $A$ and $B$ are given by

  \[ A(a) = C_1 a^{- \frac{p}{2} + \frac{1}{2}} BesselJ \left(  \frac{\sqrt{p^2 + 4 b_{44} - 2p + 1}}{4}, \frac{\sqrt{- a_{44} k }\, a^2 }{2} \right)   \]\begin{equation}
      +\, C_2 a^{- \frac{p}{2} + \frac{1}{2}} BesselY \left(  \frac{\sqrt{p^2 + 4 b_{44} - 2p + 1}}{4}, \frac{\sqrt{- a_{44} k }\, a^2 }{2} \right),
  \end{equation}
\begin{equation}
    B(y) = C_3 BesselJ \left(  2 \sqrt{ d_{44}}, 2 \sqrt{-c_{44}\, y}   \right) + \, C_4 BesselY \left(  2 \sqrt{ d_{44}}, 2 \sqrt{-c_{44}\, y}   \right),
\end{equation}
  where  $ a_{44}= \frac{M^2}{\hbar^2} ~~~, ~~b_{44}=\frac{2 M \omega^2}{\hbar^2}$, $ ~~c_{44}=\frac{2 \lambda^2 V_0}{\hbar^2},$$ ~~d_{44}=\frac{2 \lambda^2 \omega^2}{\hbar^2}$
 and  $C_i$ are integration constants. The behavior of the wave functions are shown in figure 7.\\
 \begin{figure} [thbp]
\centering
	\includegraphics[width=5.5cm]{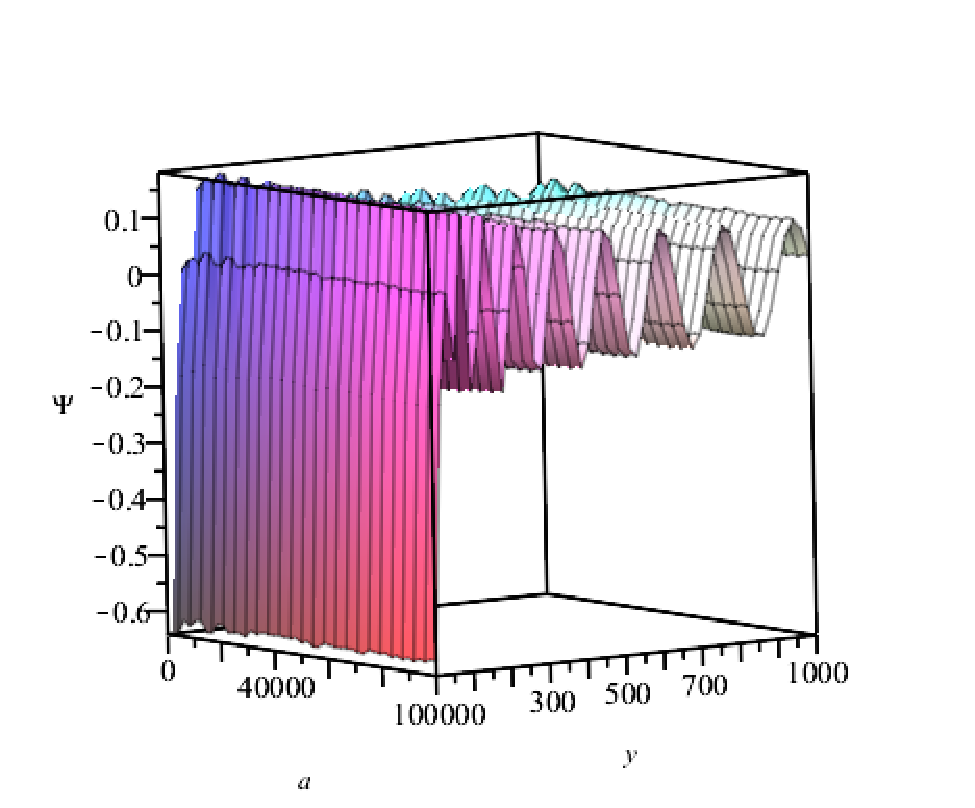}
\includegraphics[width=5.5cm]{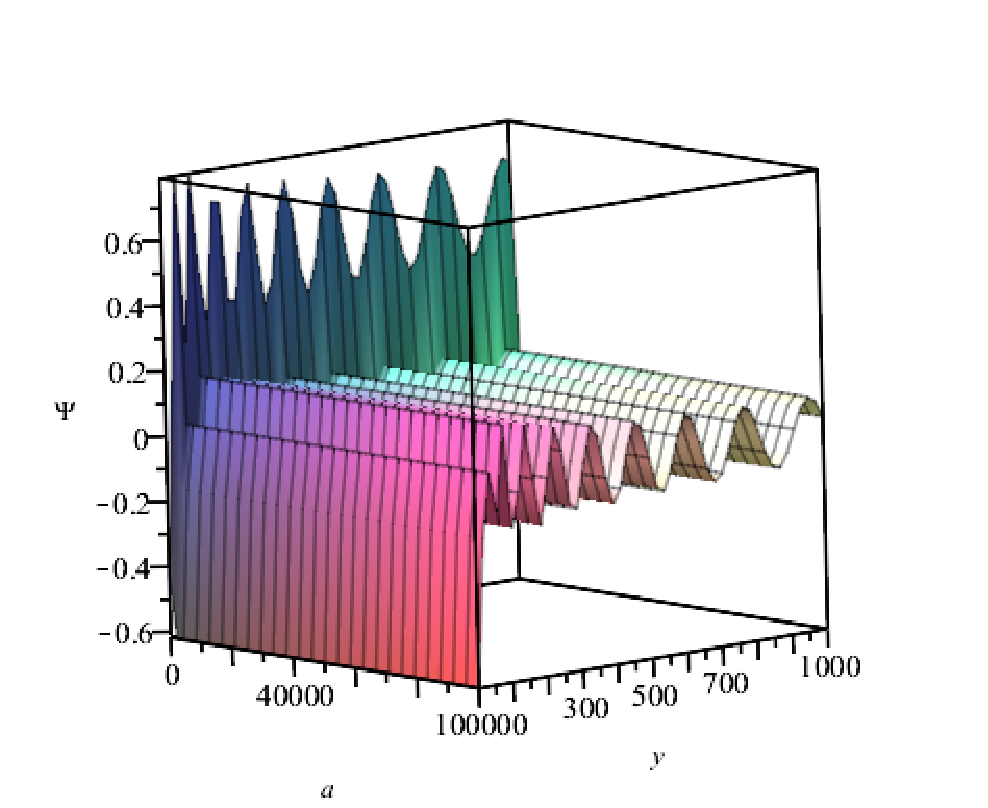}
\includegraphics[width=5.5cm]{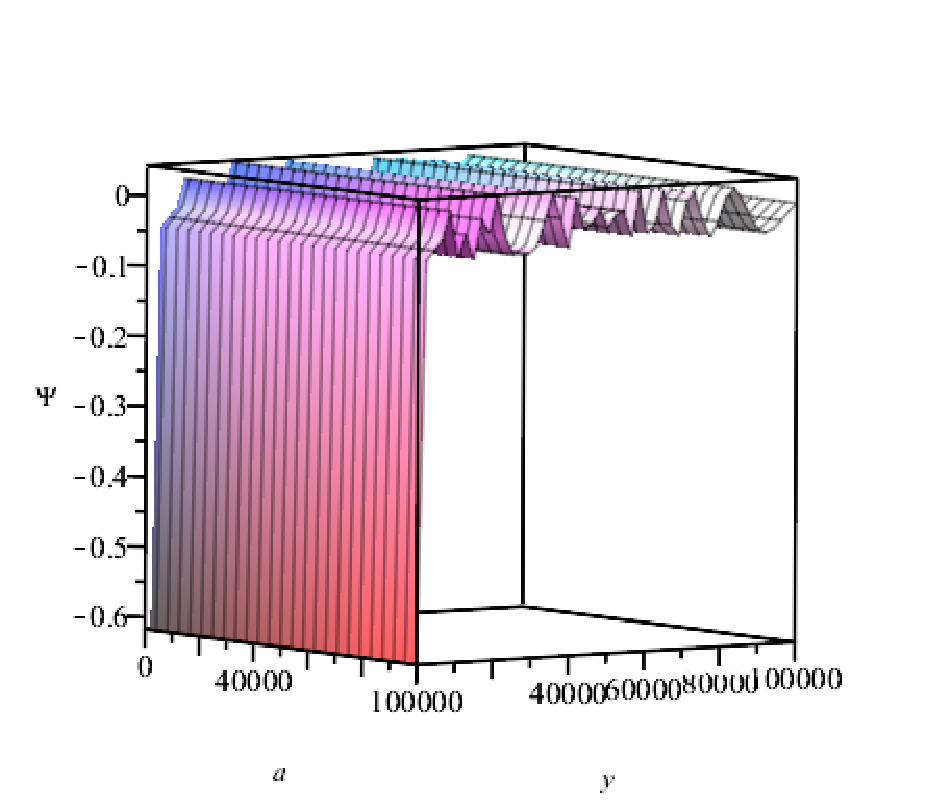}
	\caption { The wave function represents a quantum wormhole solution for different values of the parameters,   (left panel) for $ V_0=-1, p=0, k=-1$,    (middle  panel) for $ V_0=-1, p=1, k=-1$,   (right panel) for $ V_0=-1, p=2, k=-1$.   }
\end{figure}

  \section{  Quantum wormhole with perfect fluid  matter sources :}


The wave function of the universe is originated from the solution of
  Wheeler and de Witt (WdW) equation. This  is very    similar to the one dimensional
  Schrodinger's equation    without considering   time evolution.
We Consider  the universe as  homogeneous  and  isotropic  to describe the minisuperspace model with the scale factor $a (t)$  of the Friedmann-Robertson-Walker (FRW) metric as
\begin{equation}
    ds^2=-dt^2+a^2(t)\left[\frac{dr^2}{1-kr^2}+r^2(d\theta^2+\sin^2(\theta)d\phi^2)\right] .
\end{equation}
For thus FRW universe, we consider the Lagrangian as
\begin{equation}
    L=\frac{3\pi}{4G}a^3\left[\frac{\dot a^2}{a^2}-\frac{k}{a^2}+\frac{8\pi G}{3}\rho\right] .
\end{equation}
The action is defined as
\begin{equation}
    S= \int L \,dt = \int \frac{3\pi a^3}{4G}\left[\frac{\dot a^2}{a^2}-\frac{k}{a^2}+\frac{8\pi G}{3}\rho \right] \,dt .
\end{equation}
Hence the momentum conjugate  $a$ is defined by
\begin{equation}
    P=\frac{\partial L}{\partial \dot a}=\frac{3\pi}{2G}a \dot a \label{11.58}
\end{equation}



 The conservation $\partial_\mu T^{\mu \nu}=0$, where energy momentum tensor is defined by
\begin{equation*}
    T_{ab}=(\rho+\xi)u_i u_j-\xi g_{ij},
\end{equation*}
where $\rho$ is energy density and $\xi$ is the pressure, yields
\begin{equation}
    \dot \rho+3\frac{\dot a}{a}(\xi+\rho)=0 .
\end{equation}

In term conjugate momentum, one can write the Hamiltonian, $H=P\dot a-L$, as
\begin{equation}
    H(P,a)=\frac{3\pi}{4G}a^3\left[\frac{4G^2P^2}{9\pi^2a^4}+\frac{k}{a^2}-\frac{8\pi G}{3}\rho\right].
\end{equation}

We write the WdW equation as the evolution of the universe is fully described by its quantum states which satisfy the WdW equation.\\

Now, to write the most general form of WdW equation, we use the most general quantization of momentum as
$$P^2 \to -\frac{\hbar^2}{a^p}\frac{\partial}{\partial a}\left(a^p\frac{\partial}{\partial a}\right).$$

Now the WdW equation $ H \Psi=0$, where $\Psi(a)$ is the wave function of the universe as

\begin{equation}
    \left[\frac{d^2}{da^2}+\frac{p}{a}\frac{d}{da}-\frac{9\pi^2 a^2}{4\hbar^2G^2}\left(k-\frac{8\pi G}{3}\rho a^2 \right)\right]\Psi=0 .
\end{equation}

To solve this WdW equation (22), we take a transformation as
\begin{equation}
    \Psi(a)=a^{-\frac{p}{2}}y(a) .
\end{equation}

Then equation (22)  takes the form as
\begin{equation}
    \frac{d^2y}{da^2}+V(a,\rho)y=0,
\end{equation}
where the effective potential $V(a,\rho)$ is given by
\begin{equation}
    V(a,\rho)=-\frac{9\pi^2k a^2}{4\hbar^2 G^2}+\frac{6\pi^3}{G \hbar^2}\rho a^4-\frac{(p^2-2p) }{4}\frac{1}{a^2}.
\end{equation}

The  equation  (23)  involving    energy density $\rho$,  therefore,  we  consider  some
particular types of fluid to achieve wormhole solution.\\

\textbf{type-1:}\\

Using the linear  equation of state $\xi = w \rho -b $, ( $b, w$ are constants ) the conservation equation (29) yields
\begin{equation}
     \rho =\frac{b}{1+w} +\rho_0 a^{-3(1+w)} ~~~~(\rho_0 ~is~integration~ constant).
\end{equation}
With this value of $\rho$,  WdW equation (23) assumes the form as
\begin{equation}
\frac{d^2y}{da^2}+\left(-\frac{9\pi^2k }{4 \hbar^2G^2}a^2+\frac{6 b   \pi^3}{G (1+w)\hbar^2}a^4+\frac{6 \rho_0 \pi^3 }{G \hbar^2} a^{(1-3w)}-\frac{(p^2-2p) }{4}\frac{1}{a^2} \right)y=0.
\end{equation}
This complicated equation   involving $a$ is difficult to solve,  therefore, it convenient to explore  the behaviors of the wave function $\psi$ near $a \approx 0$ and $a \rightarrow \infty$. In the limit $a \rightarrow 0$, the term $a^{-2 }$ is dominating whereas   the dominating term  is ~ $a^4$   in the limit $a \rightarrow \infty$ for $-1<w<1$.  Now the equation (36) takes the following forms
\[     \frac{d^2 y}{\partial a^2}  +\frac{p^2-2p}{4a^2} =0,\]
\[   \frac{d^2 y}{\partial a^2}+  +b_{44} a^4 =0.\]
We obtain the following solutions of  these equations as
\begin{equation}y(a) = C_1a^{1 - \frac{p}{2}} + C_2a^{\frac{p}{2}},\end{equation}
\begin{equation}y(a) = C_3\sqrt{a} BesselJ\left(\frac{1}{6}, \frac{a^3}{3}\right) + C_4\sqrt{a} BesselY\left(\frac{1}{6}, \frac{a^3}{3}\right),\end{equation}
where $C_i$ are integration constants. \\

 \begin{figure} [thbp]
\centering
	\includegraphics[width=5.5cm]{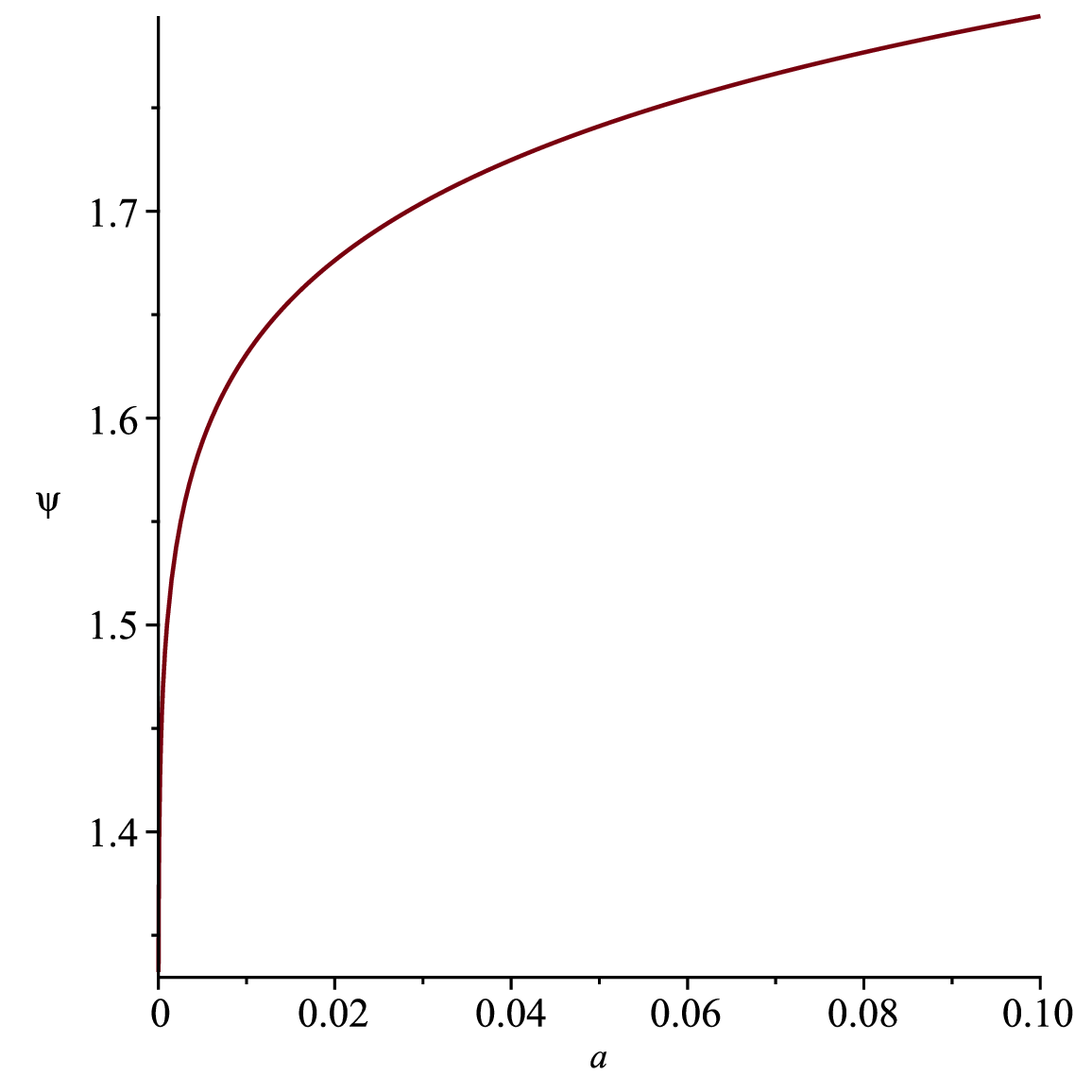}
\includegraphics[width=5.5cm]{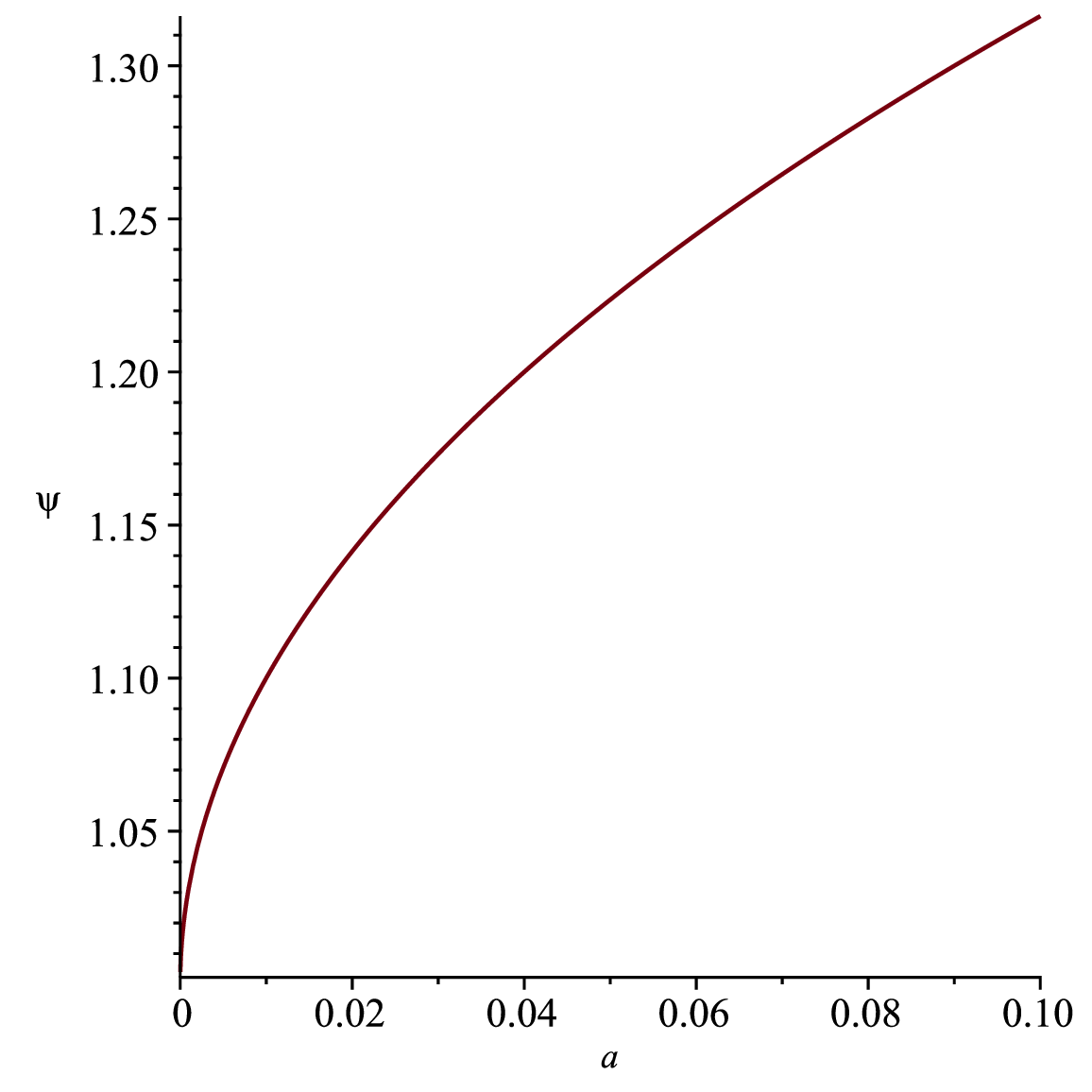}
\includegraphics[width=5.5cm]{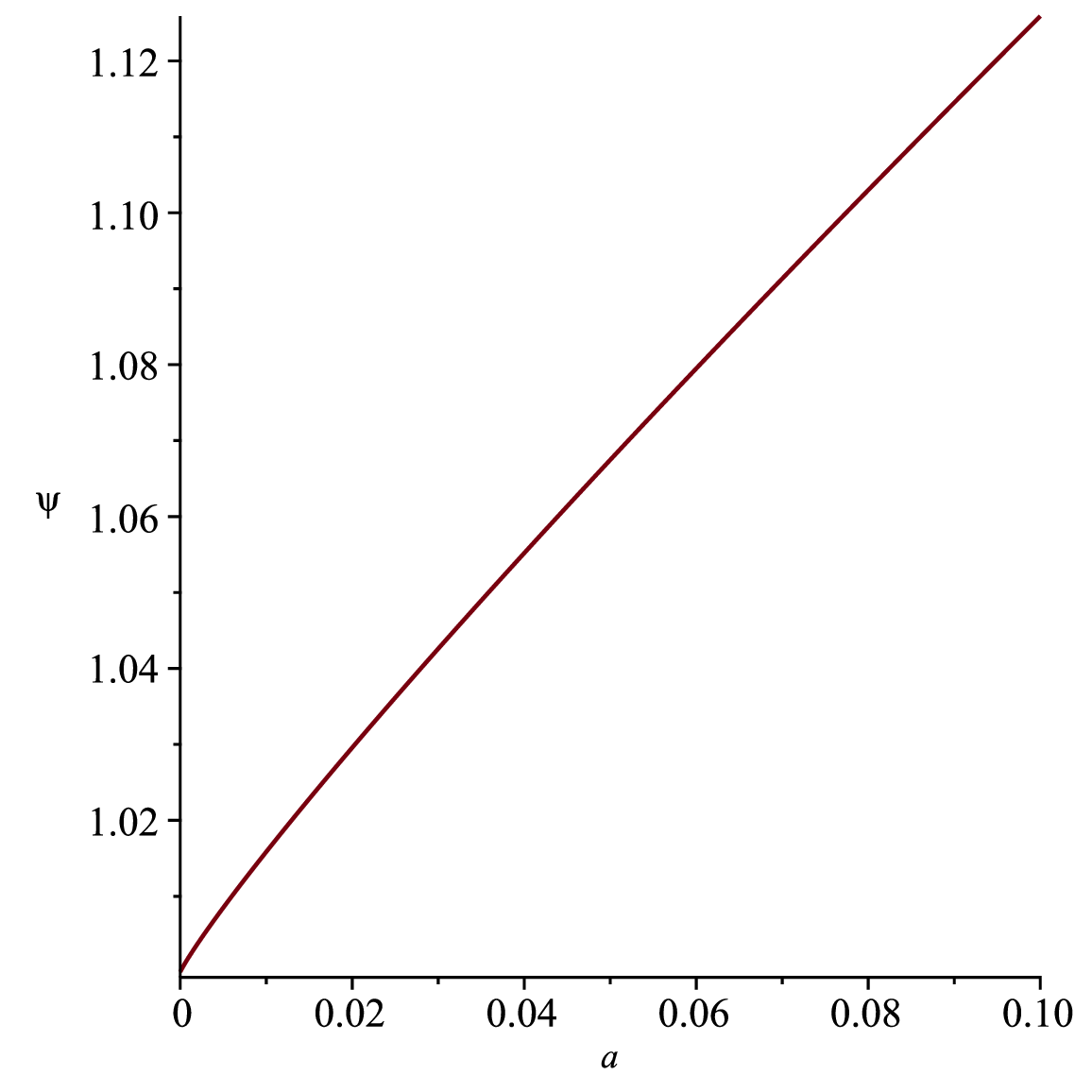}
	\caption { The wave function for scale factor $a \approx 0$ represents the quantum wormhole solution (left panel) for $p=.9$,  left panel) for $p=.5$     (right panel) for $p=.1$ .   }
\end{figure}
 \begin{figure} [thbp]
\centering
	\includegraphics[width=5.5cm]{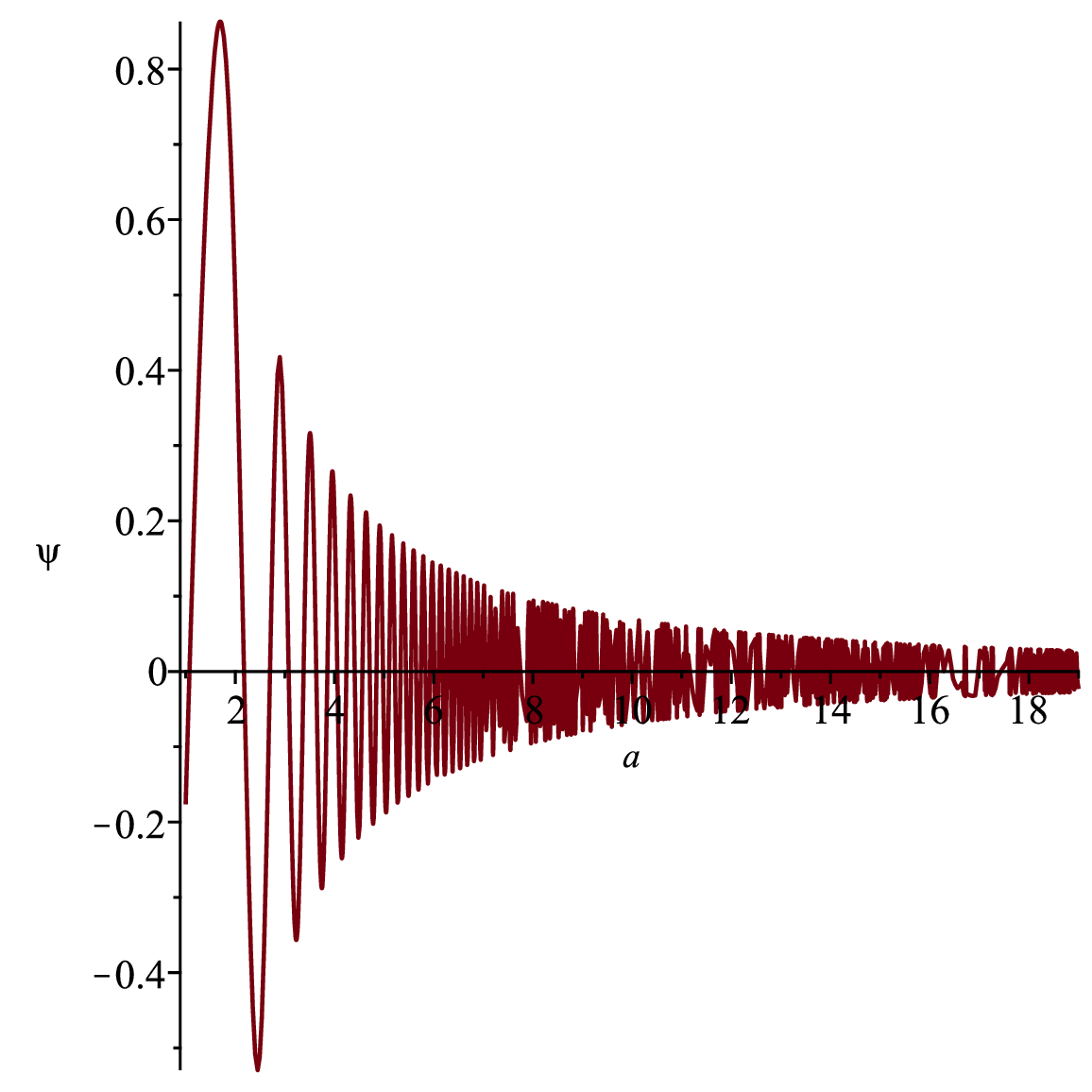}
\includegraphics[width=5.5cm]{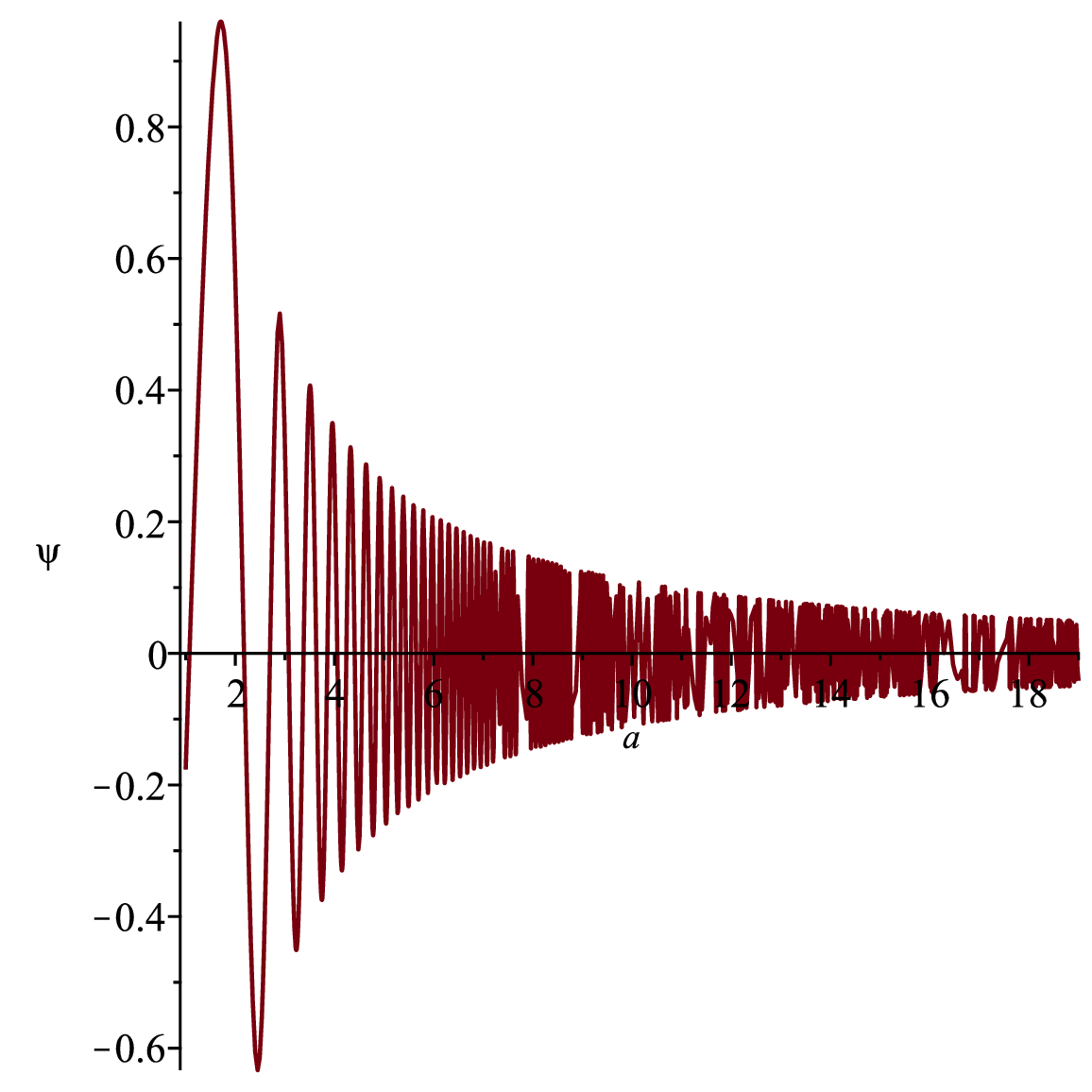}
\includegraphics[width=5.5cm]{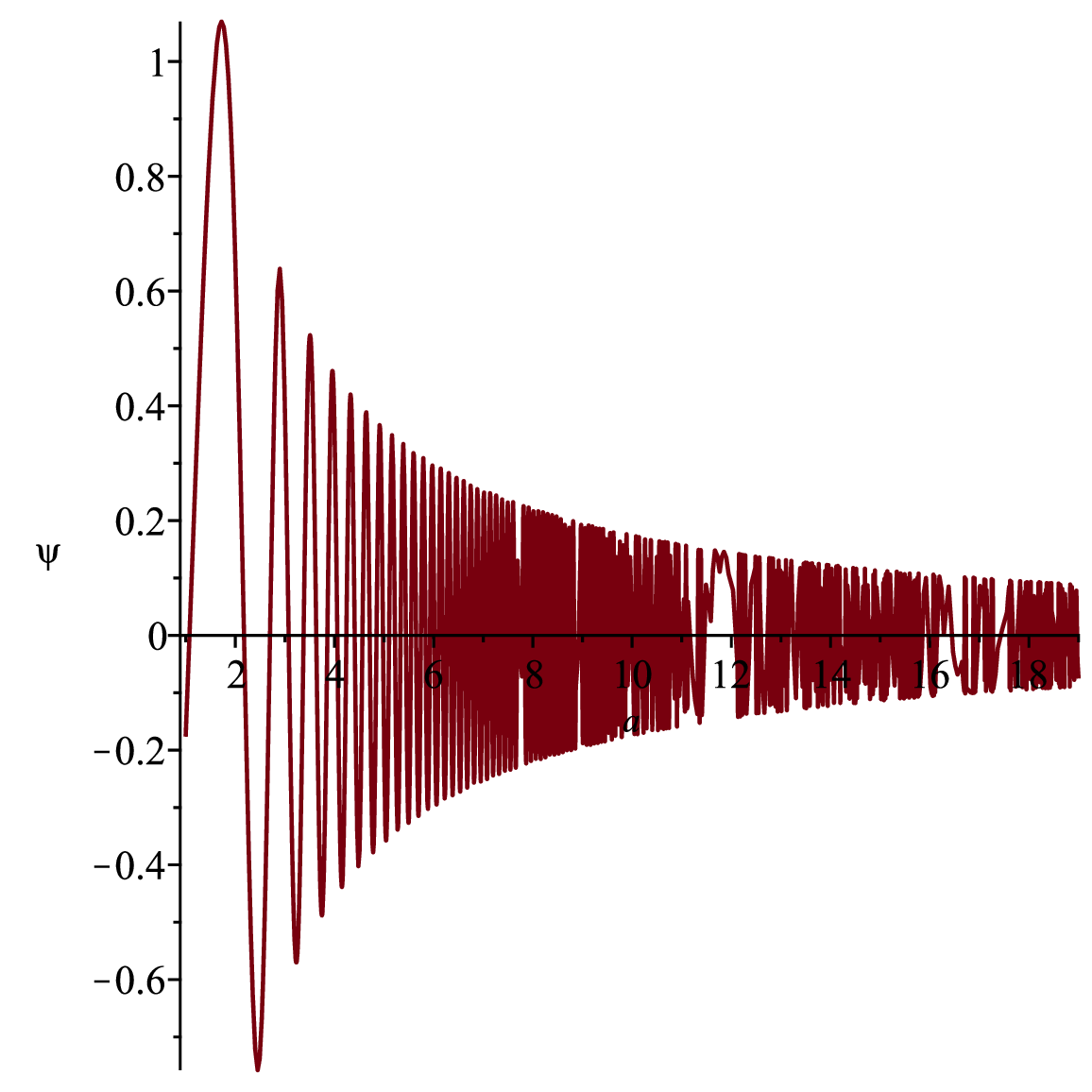}
	\caption { The wave function for large value of scale factor i.e. $a \rightarrow \infty$,  represents the quantum wormhole solution for $b_{55} =1$ (left panel) for $p=.9$,  left panel) for $p=.5$     (right panel) for $p=.1$.   }
\end{figure}

\textbf{type-2:}\\

Matter filling universe is mixture of a radiation and a non-relativistic component.\\

Therefore
\begin{equation}
    \rho  = \rho _{mo} \left( \frac{a_0}{a} \right) ^3 + \rho _{ro} \left(\frac{a_0}{a} \right) ^4,~~( \rho _{mo} ~and  ~\rho _{ro} ~are~ constants).
\end{equation}

With this value of $\rho$,  WdW equation (33) assumes the form as
\begin{equation}
\frac{d^2y}{da^2}+\left(-\frac{9\pi^2k }{4 \hbar^2G^2}a^2+\frac{6  \rho _{mo}   \pi^3}{G (1+w)\hbar^2}a+\frac{6  \rho _{ro}   \pi^3}{G (1+w)\hbar^2}-\frac{(p^2-2p) }{4}\frac{1}{a^2}\right)y =0.
\end{equation}

As before, we explore the behaviors of the wave function $\psi$ near $a \approx 0$ and $a \rightarrow \infty$. For these conditions, equation (40) can be written as
\[     \frac{d^2 y}{\partial a^2}  +\frac{p^2-2p}{4a^2} =0,\]
\[   \frac{d^2 y}{\partial a^2}-k b_{66} a^4 =0.\]
The  solutions of  these equations are given by
\begin{equation}y(a) = C_1a^{1 - \frac{p}{2}} + C_2a^{\frac{p}{2}},\end{equation}
\begin{equation}y(a) = C_3\sqrt{a} BesselJ\left(\frac{1}{4}, \frac{a^2}{2}\right) + C_4\sqrt{a} BesselY\left(\frac{1}{4}, \frac{a^2}{2}\right),\end{equation}
where $C_i$ are integration constants.
\begin{figure} [thbp]
\centering
	\includegraphics[width=5.5cm]{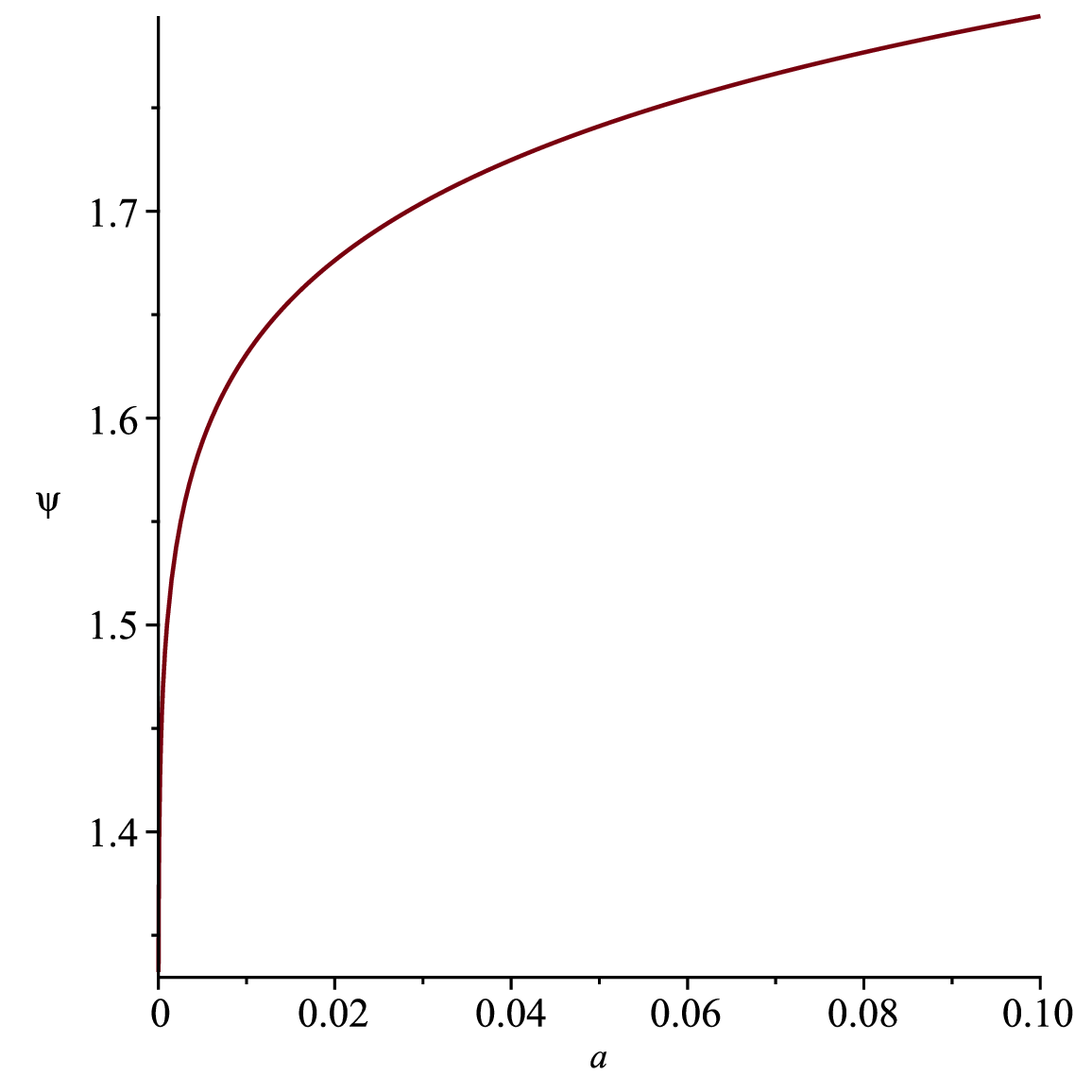}
\includegraphics[width=5.5cm]{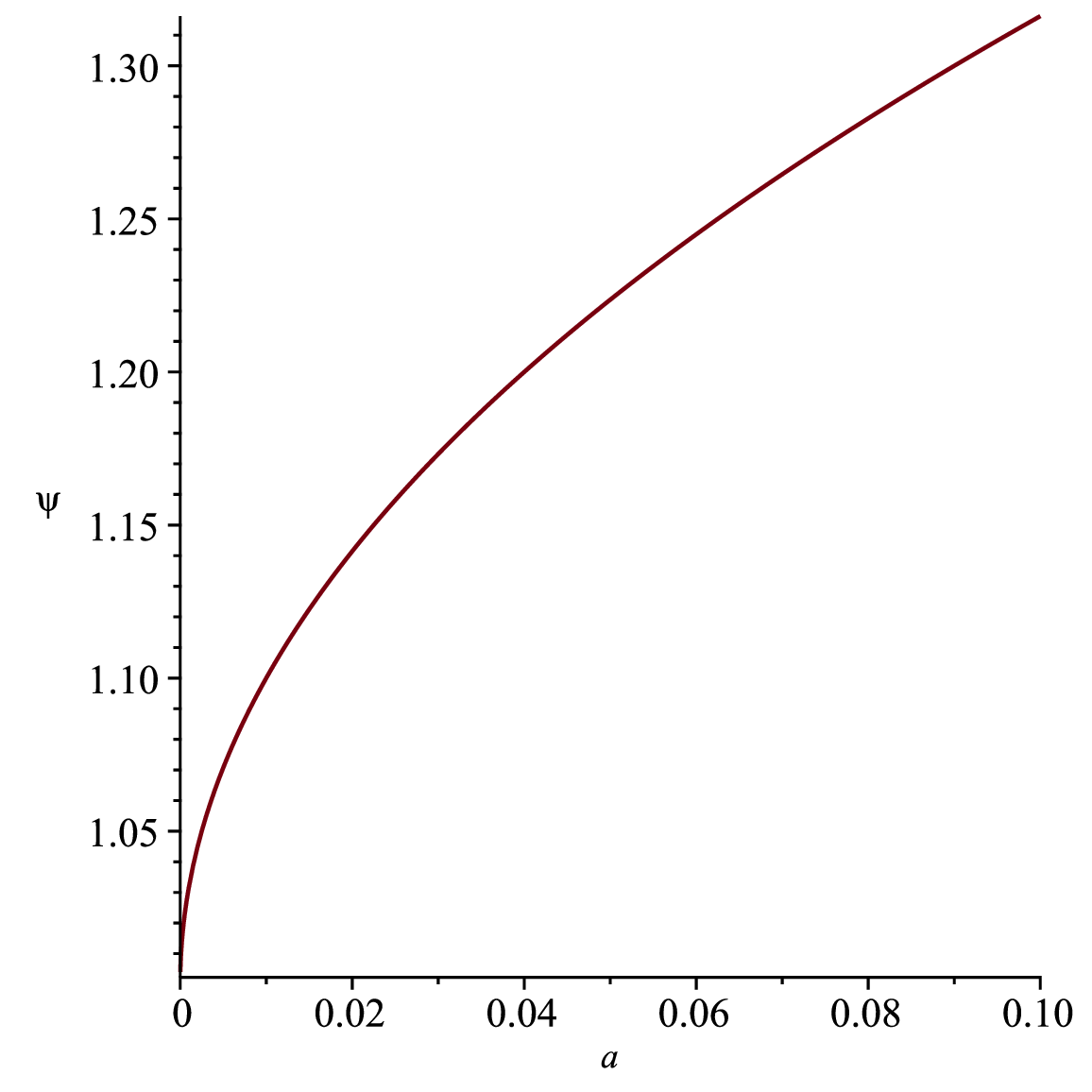}
\includegraphics[width=5.5cm]{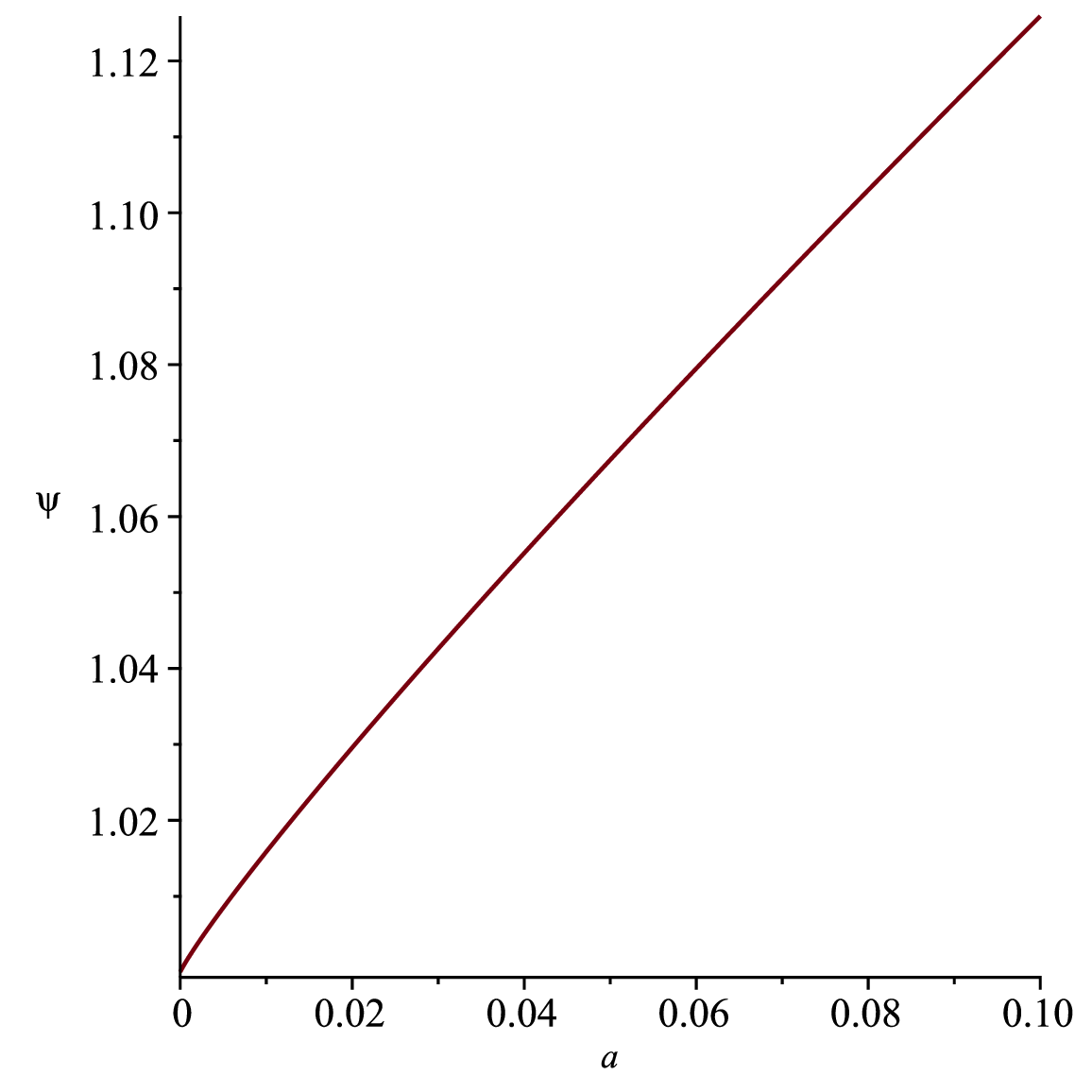}
	\caption { The wave function for scale factor $a \approx 0$ represents the quantum wormhole solution (left panel) for $p=.9$,  left panel) for $p=.5$     (right panel) for $p=.1$ .   }
\end{figure}
 \begin{figure} [thbp]
\centering
	\includegraphics[width=5.5cm]{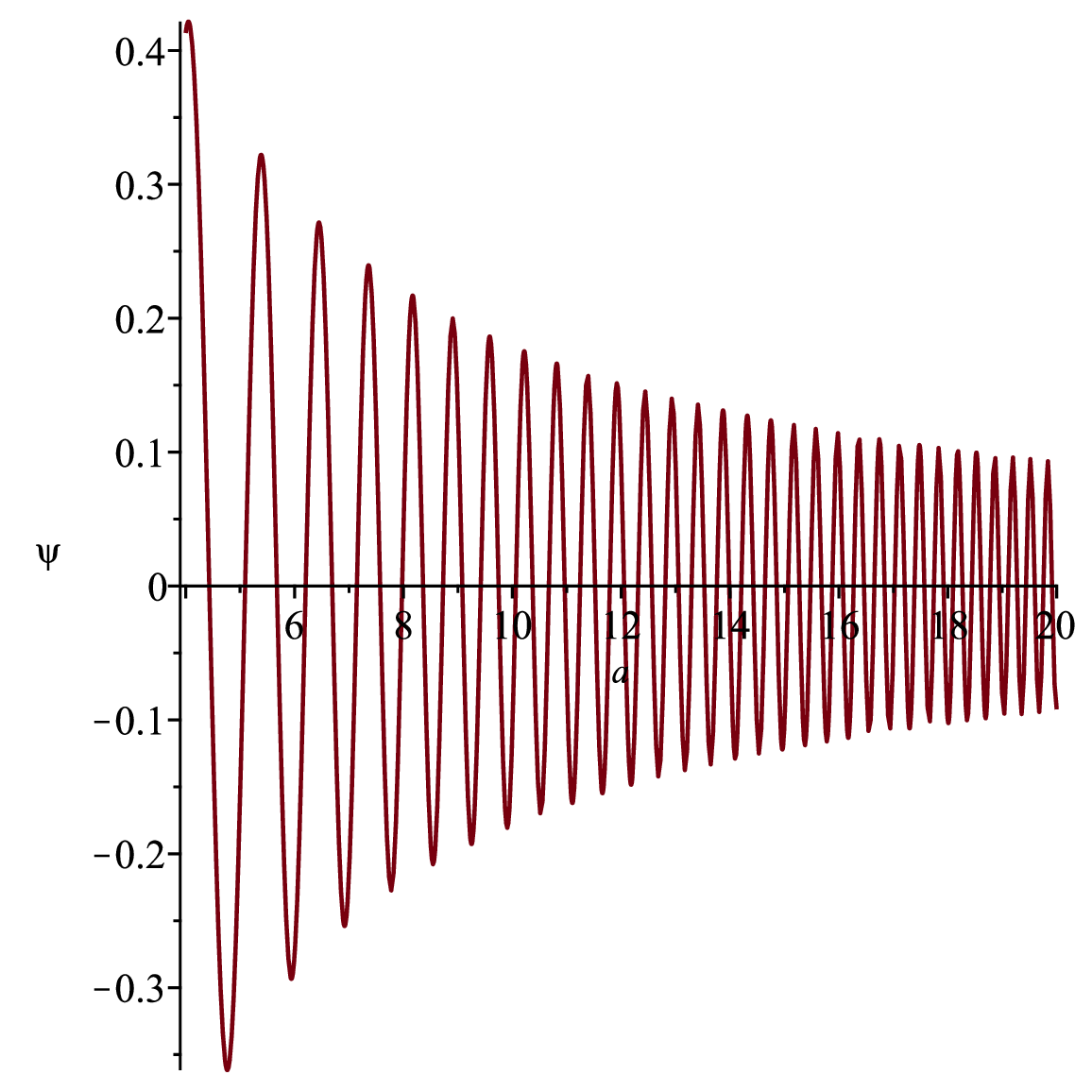}
\includegraphics[width=5.5cm]{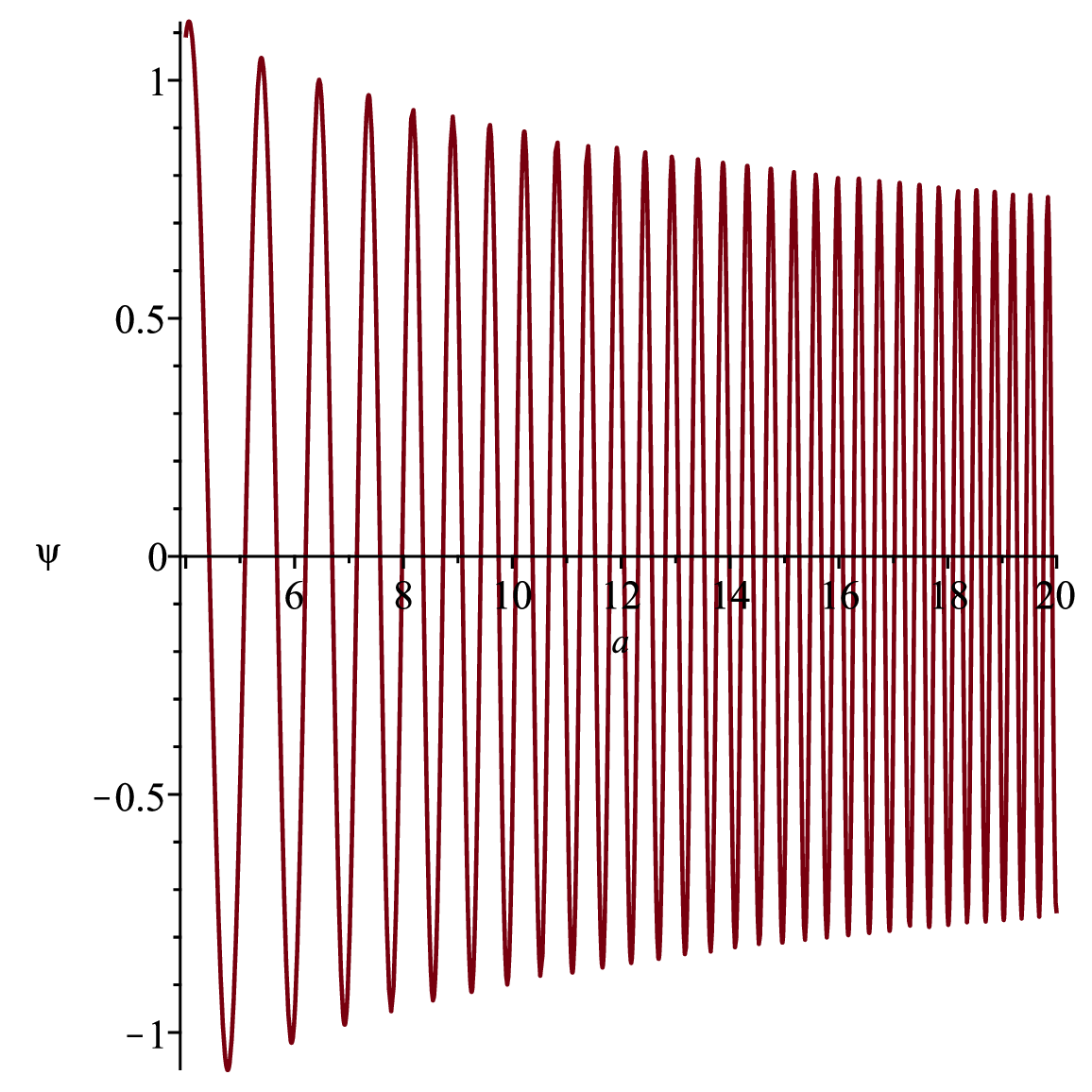}
\includegraphics[width=5.5cm]{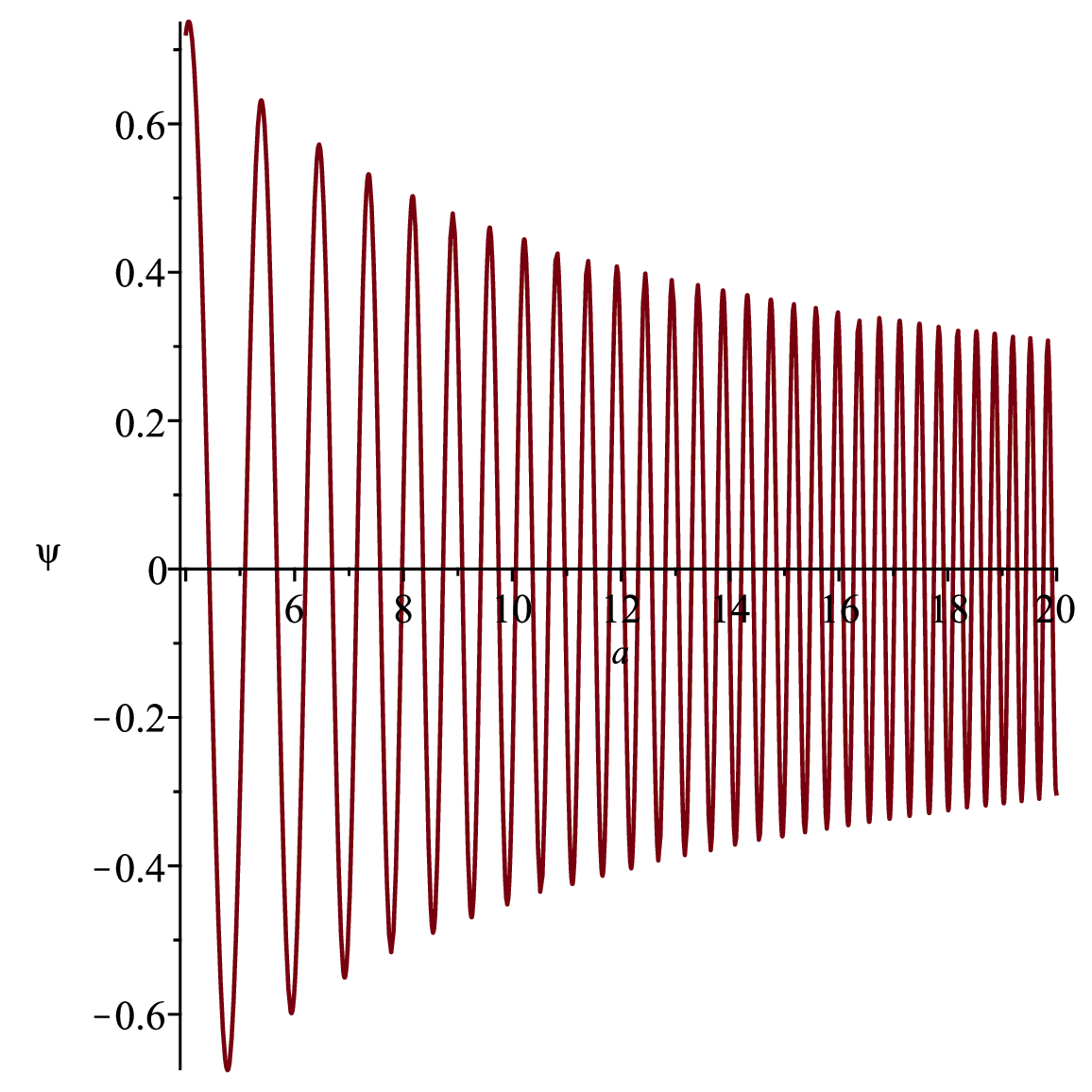}
	\caption { The wave function for large value of scale factor i.e. $a \rightarrow \infty$,  represents the quantum wormhole solution for $k =-1,  a_{66} =1$ (left panel) for $p=.9$,  ( middle panel) for $p=.5$     (right panel) for $p=.1$.   }
\end{figure}

 \section{concluding remarks:}

In this paper,  we have explored the Euclidean quantum wormholes in the context of the FRW Universe minisuperspace model.
To explore microscopic wormholes, one needs to investigate the early universe extensively, characterized by either vacuum dominance or the possible presence of a scalar field. The application of the Hawking-Page boundary condition proves valuable in identifying these minuscule wormholes.
Initially we have considered minimally coupled scalar field coupled to gravity and taken its action in order to determine the Hamiltonian constraint equation which leads to the Wheeler-DeWitt equation and in the latter part we have considered quantum wormholes with homogenous and isotropic perfect fluid matter sources achieving ultimately the WDW equation. We have followed the general quantization of momenta in the Dirac quantization procedure to achieve the WDW equation. From the case studies we have inspected certain wormhole solutions and found cases where the criteria of the wave function to represent wormholes is satisfied and also we came across cases where it doesn't adhere to wormholes. We have graphically explored the aspects of the wave function plotted against the scale factor, and in the former case against the scalar field as well. Hence, within minisuperspace models featuring both minimally coupled scalar field and perfect fluid matter sources, there exist, or at least seem to exist, infinite discrete spectra of solutions to the Wheeler-DeWitt equation. These solutions exhibit regularity throughout, encompassing points such as $a \rightarrow 0$   and display exponential damping at $a\rightarrow \infty$. These solutions can be interpreted as wave functions corresponding to wormholes.

\subsection{Acknowledgments}
FR    would like to thank the authorities of the Inter-University Centre for Astronomy and Astrophysics, Pune, India for providing the research facilities.  BSC   and  AI  thank  UGC, Govt. of India for financial support in terms of a fellowship. FR   is  also   thankful to SERB, DST and   DST FIST programme     (  SR/FST/MS-II/2021/101(C))   for financial support respectively.

\end{document}